# Drying behavior of dense refractory ceramic castables. Part 1 – General aspects and experimental techniques used to assess water removal

A. P. Luz[(1),*], M. H. Moreira[(1)], M. A. L. Braulio[(2)], C. Parr[(3)], V. C. Pandolfelli[(1)]

[(1)] Federal University of São Carlos, Graduate Program in Materials Science and Engineering (PPGCEM), Rod. Washington Luiz, km 235, São Carlos, SP, 13565-905, Brazil.

[(2)] 4Cast, Technical Assistance and Consultancy on Refractories,
Rua Aristides de Santi, n° 06, Bairro Azulville I, 13.571-150, São Carlos, SP

[(3)] Imerys Aluminates SA, Paris La Defense, 92800, Puteaux, France.

*Corresponding author at: t*el.:* +55-16-3351-8601
E-mail: anapaula.light@gmail.com or analuz@ufscar.br

**Abstract**

Despite the continuous evolution on the performance of refractory ceramic products, monolithic materials still require special attention during their processing steps as various phase transformations may take place during the curing, drying and firing stages. Drying is usually the longest and the most critical process observed during the first heating cycle of refractory linings, as the enhanced particle packing and reduced permeability of the resulting microstructure may lead to recurrent explosive spalling and mechanical damage associated with dewatering and the development of high steam pressure at the inner regions of such dense materials. In this context, this review article mainly addresses *(i)* the theoretical aspects related to the drying process of dense refractories, *(ii)* the influence of the phase transformations derived from the binder additives, and *(iii)* the usual and advanced experimental techniques to assess the water removal from consolidated castable pieces. Many studies have pointed out that due to the complex nature of this phenomenon (i.e., considering combined thermal stresses and pore pressure, heterogeneous microstructure, evolving pore structure with temperature, etc.), the mechanisms behind the water withdrawal and castables' explosive spalling are lacking further understanding and, consequently, it has been difficult to save time and energy during the first heating of industrial equipment lined with ceramic materials. On the other hand, different methods are used for refractory spalling assessment and many efforts have been carried out in applying *in situ* imaging techniques (such as NMR and neutron tomography) to follow the moisture evolution during such thermal treatments. These novel techniques, also addressed in this review, might be of particular importance to provide more accurate data for the validation of many state-of-the-art numerical models, which can be used to predict the steam pressure developed in refractory systems and help in the design of proper heating schedules for such products.

## 1. Introduction

Processing of refractory castables is usually based on a mixing stage (which requires adding water or some other liquid to provide suitable molding of the pieces), followed by the curing period (where the hardening of the fresh mixture takes place) and a further drying process to withdraw the added liquid from the ceramic microstructure before firing. Drying has become a huge challenge to refractory producers and end-users, because there is a growing demand for shorter equipment maintenance time, which triggers the development of faster drying materials. Nevertheless, the advances in the design of refractory castables over recent decades resulted in products with enhanced particle packing and reduced permeability, requiring even longer heating schedules, as the ones previously used were no longer feasible due to the recurrent explosive spalling and mechanical damage observed during dewatering [1–3].

Among all castable processing steps, dry out is usually the longest one, which leads to three important issues: *(i)* high energy consumption is required, *(ii)* its impact on the production halt (lost income), and *(iii)* the mechanical damage derived from the dewatering step. The latter, results in shorter working life for the ceramic lining and extra spending on maintenance and material [1,4]. As reported by some authors [5,6], the selection of proper heating rates and dwell times is mostly based on empirical knowledge and conservative drying schedules are commonly preferred. Consequently, the dry out stage of dense castables is an expensive and time-consuming procedure (including heating rates lower than 30°C/hour and many dwell times at intermediate temperatures), as it delays resuming the operational conditions [5].

Therefore, this subject has motivated many studies focused on better understanding the water distribution and advance of the drying front in green ceramic bodies [2,7–10], as well as the development of numerical models of heat and mass transfers in moistened ceramics to predict safe and optimized heating schedules for refractory compositions [11–14].

Considering these aspects, this review article addresses the following subjects: *(i)* the theoretical aspects related to the drying process of dense refractories, *(ii)* the influence of the phase transformations derived from the binder additives, and *(iii)* the main experimental techniques used to assess the water removal from consolidated castable pieces. A forthcoming publication (Part 2) should complement the present paper, pointing out *(i)* the main drying agents and how they affect the resulting refractories' microstructure, and *(ii)* design of drying schedules for different refractory compositions.

## 2. General aspects of the drying process of dense refractory castables

Usually, adding water to refractory products is required due to two main aspects: *(i)* to induce a more efficient mixture and homogenization of the raw materials, making it easier to convey the fresh castable and shape it, and *(ii)* to favor the development of chemical reactions and further precipitation of hydrated phases due to the interaction of the liquid with hydraulic binders (i.e., calcium aluminate cement = CAC, hydratable alumina = HA, magnesia) contained in the products' formulation [3,15,16]. Although some refractories may be prepared without water or with other liquid components (i.e., phenolic resins [17]), $H_2O$ is cheaper, more practical and does not present environmental issues, such as the release of toxic volatile species commonly observed for organic binders. Hence, adding water during the processing of most commercial castables cannot be avoided and this is an important component of the available refractory systems.

## 2.1. Main drying stages

*Drying* can be defined as the withdrawal of water from a material, driven by pressure, concentration and/or temperature gradients, involving both heat transfer (by conduction, convection or radiation) and mass movement (by diffusion or capillary) [1,18]. In the case of dense castables, the water release takes place mainly through the open porosity contained in the matrix region (fraction of the refractory microstructure comprised by fine components and the reaction products derived from the binder interaction with the liquid, Fig. 1) and the permeable interfaces between matrix and aggregates (coarse raw materials), resulting from the packing flaws associated with the differences in the particles size [15].

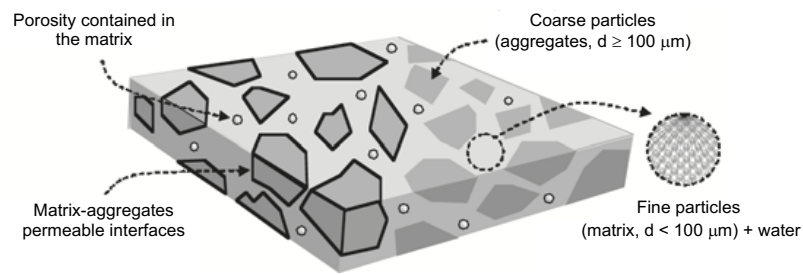

Figure 1: Sketch of refractory castable structure pointing out its components and permeable paths that help the drying process (adapted from [15]).

### 2.1.1. Evaporation, ebullition and hydrate decomposition

After the castables' mixing and curing steps, a continuous liquid film is found surrounding the surface and filling in the formed structure, as indicated in Fig. 2a. Water should be initially withdrawn from the region just beneath the surface, without any change in pressure as the temperature is usually constant at the beginning of the drying process. The vaporization stage should take place slowly and it depends on temperature and humidity available in the environment. After removing this external liquid film, solid/air interfaces are generated, and meniscuses tend to be formed among the particles (due to the water surface tension) to keep the liquid cohesion. This condition induces the development of strong capillary forces that pull the water from the inner region of the refractory to its surface, and it should continue as long as the balance between water dragging from deeper areas and its release to the atmosphere is constant (Fig. 2b). Consequently, the drying rate during this period is steady and usually called the *constant-rate period* (CRP) [1,15].

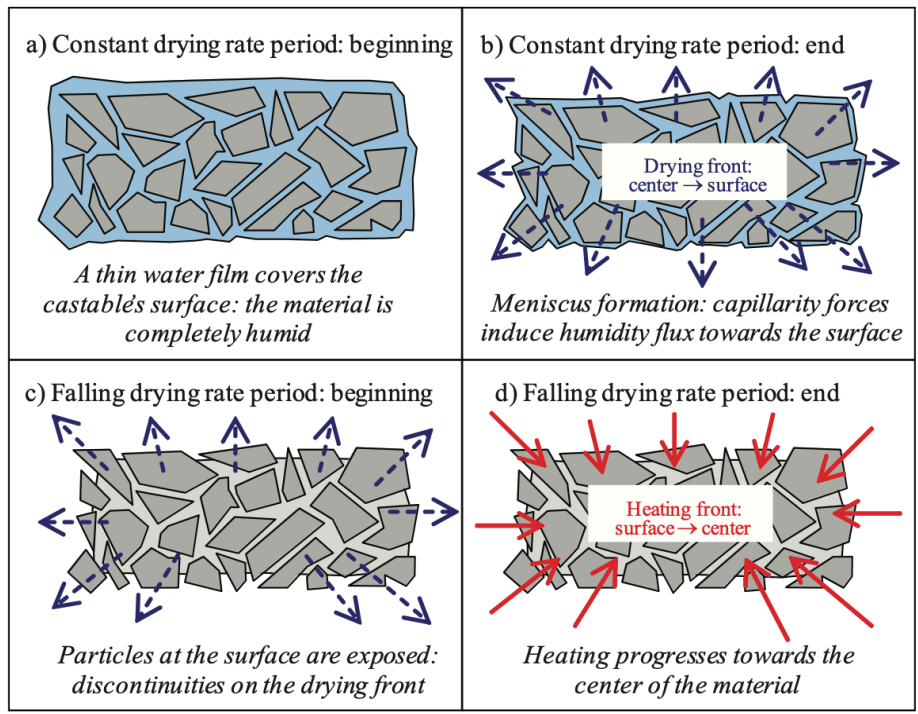

Figure 2: Initial castable's drying stages in the presence of a liquid film of water (adapted from [1,15]).

However, the dry out of castables is commonly carried out under continuous heating, which leads to a marked vapor pressure increase when the temperature is around 100°C. Thus, at this stage, a discontinuity is observed between the deeper and humid layers of the structure and the dried surface, as the surface dewatering rate is higher than the water migration speed from the inner regions of the dense ceramic (Fig. 2c). After this, water transfer will rely on the vapor diffusion through the permeable paths available in the microstructure. Due to this feature, the drying rate decreases, giving rise to the so-called *falling-rate period (FRP)*. Besides that, during the vaporization step, the temperature and heating rate at the surface of the material are lower than those of the environment as a consequence of the higher heat capacity of the water when compared to the solid components. This aspect plays an important role at the final stages of the drying process, as less $H_2O$ is present among the solid particles, and hence a higher amount of heat can be absorbed by the refractory (Fig. 2d). Consequently, in order to recover the thermal equilibrium, the partially dried surface of the ceramic lining usually faces a greater heating rate than the one heading the environment (i.e., furnaces or drying burners) [15].

Fig. 3 highlights the drying process for monolithic compositions as a function of heat transfer and water movement. During the continuous heating of such materials, the *drying front* is located at the point where an inflection is observed in the temperature profile along the refractory thickness. Another important aspect is the total amount of water contained in the castable, which comprises the sum of the funicular and pendular ones, related to the *CRP* and *FRP* stages, respectively [19].

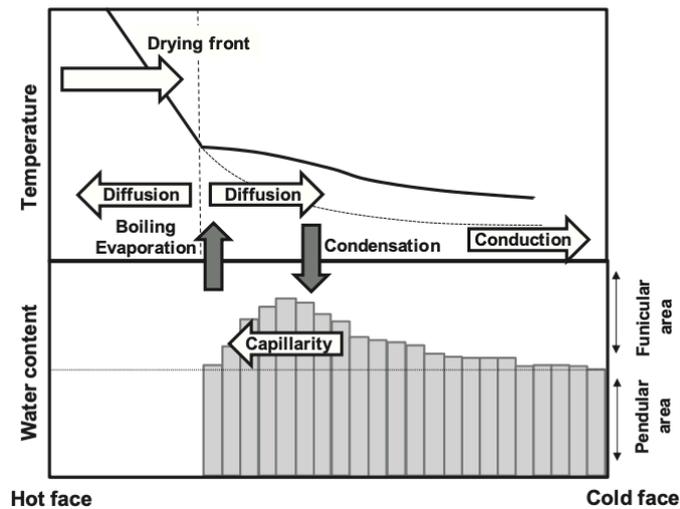

Figure 3: Sketch of the heat and mass (water) transfer throughout the castable's lining during drying [19].

Based on Fig. 3 and the presented drying front location, no free water was contained in the refractory's hot face, indicating that the process has finished in this area. However, as heat is transferred by conduction to deeper regions of the structure, water starts boiling ahead of the drying front and a fraction of the formed vapor diffuses to the heating side and is withdrawn from the material. On the other hand, the other vapor fraction tends to move to the cold face and condensate, increasing the likelihood of heat transfer by thermal conduction to this relatively low temperature area. As this process progresses, the drying front is continuously shifted to the right side (deeper region of the material), up to the point where water is fully removed from the refractory [19].

In the case of castable's formulations containing hydraulic binders (i.e., CAC, HA or MgO), the dry out process usually takes place up to 550-600°C (depending on the material's permeability and the applied heating rates) and it can be divided into three different stages: *(i)* evaporation, *(ii)* ebullition and *(iii)* dehydration (hydrate decomposition) [5,15,20,21]. The relative amount of physical (free water expelled during evaporation or ebullition) and chemically bonded (hydrates) $H_2O$ depends on the total liquid quantity added for mixing, as well as the hydraulic binder content.

Water evaporation and ebullition, as well as hydrate decomposition can be recorded as a function of temperature and time by using thermogravimetric measurements of refractory samples. As pointed out in Fig. 4a, when the castable's temperature reaches 100°C, the ebullition starts and leads to a higher mass loss rate than that observed during evaporation, as it is then governed by the water vapor pressure instead of the environmental aspects (temperature and humidity, as previously seen for evaporation). This is the most critical dewatering step and where most likely spalling takes place for low and ultra-low castables containing calcium aluminate cement as a binder. As the vapor is quickly formed and usually located at a certain depth from the surface (generating clumps of pressurized steam), it is not easy to attain a suitable and accurate balance between the amount of gas generated inside the body and its withdrawal at the surface (which is dependent on the heating rate, castable's permeability and thickness) [15,22]. Such features are responsible for the high explosion risk presented by refractory monolithics during heating.

Moreover, Fig. 4b shows that the temperature and heating rate on the external surface and central region of the refractory specimens are lower than that set up for the furnace equipment, during evaporation and most of the ebullition stage (up to 160°C). This behavior is due to the fact that H$_2$O vaporization is a first-order transformation with high latent heat, which requires more energy to warm up the water than the solid components. Thus, dewatering is strongly affected by the refractory's temperature gradient and it is expected that a variation of vapor pressure along the material's thickness should be detected [1,15,21].

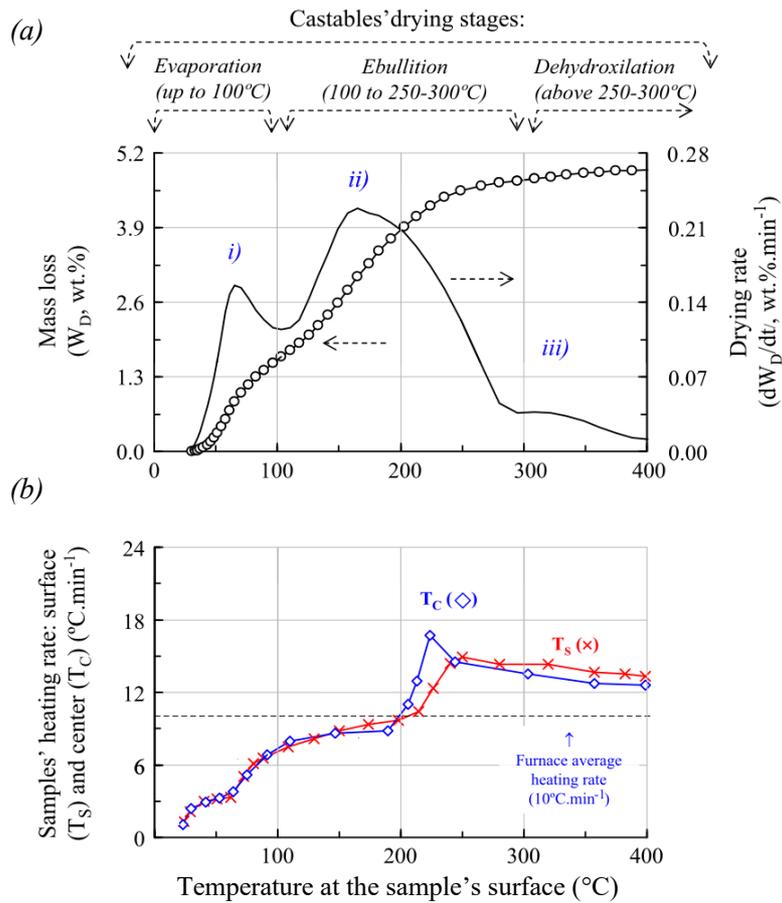

Figure 4: Drying behavior evaluation based on (a) thermogravimetric tests and (b) internal temperature profile at the sample's surface or center. The evaluated composition comprised an alumina-based castable bonded with 2 wt.% of calcium aluminate cement [21].

The reduction in the mass loss rate (Fig. 4a) is an alternative to identify the end of the ebullition and beginning of the last drying step, which is the hydrate decomposition. In general, this third stage is detected in the 180 to 600°C range, depending on the binder source, its added content and the curing conditions used while preparing the refractory [23–26]. Due to the important role of this component in the overall performance of the castable compositions, Section 2.2 will focus on pointing out the main available binders, as well as their influence on the resulting microstructure during the curing and drying steps.

### 2.1.2. Spalling mechanisms

It is important to highlight that there are two drivers for the castables' spalling: thermal strain caused by rapid heating and internal pressures due to the water removal [10,27]. As the temperature at the surface starts to move inwards a number of processes take place. Initially, the humid refractory will expand as temperature increases, then progressively its permeability will also raise due to thermal structural changes (i.e., hydrate decomposition, polymeric fibers decomposition, etc.). Eventually the temperature is high enough and water starts to evaporate and move through the pores towards the heated surface due to pressure and humidity differences. However, the water vapor movement might be limited due to the reduced permeability of the microstructure. This resistance to gas flow means that pressure will be developed within the castable body and a zone where boiling occurs will emerge [10].

According to Antoine's equation (which is valid in the range of 0 to 374°C for liquid-vapor equilibrium condition [28]), the water vapor pressure ($P_v$) follows an exponential relationship with the temperature and is described for a closed liquid/vapor aqueous system by:

$$P_v = exp\left(A - \frac{B}{T+C}\right) \qquad (1)$$

where $P_v$ is given in Pascal, $T$ is the temperature (in Kelvin), and *A, B, C* are empirical dimensionless constants (for water, A = 23.33; B = 3841.22; C = -45.00).

Fig. 5 highlights the water vapor pressure values, as calculated by Eq. 1, and the typical green mechanical strength of castables obtained via splitting tensile strength measurements [15,21]. Depending on the refractory system, boiling in the 160°C range is already dangerous and the resulting $P_v$ (between 0.6 - 2.3 MPa as shown in Fig. 5) can be high enough to induce the explosion of this material [29].

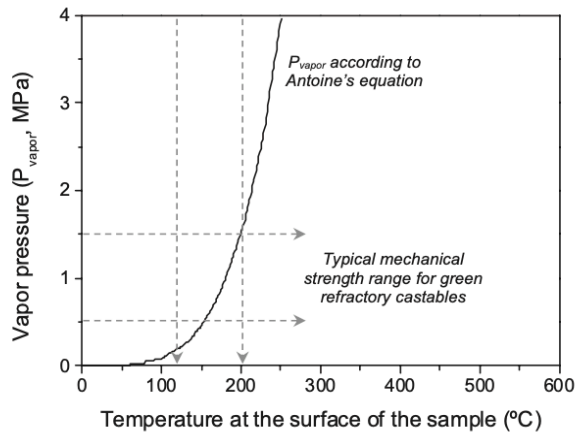

Figure 5: Vapor pressure ($P_v$) exponential increase with the temperature according to Antoine's equation (adapted from [15]).

As the imposed pressure gives rise to tri-axial tensile stress, the required condition for the castable's rupture can be easily attained, leading to spalling (Fig. 6b). On the other hand, if the formed vapor is released by the permeable paths contained in the microstructure, the ebullition stage can end up at temperatures ranging from 250 to 300°C without spalling (Fig. 6a). The complexity associated with the design of drying schedules for castables is also related to the type and amount of the formed hydrated phases, as they might change depending on the selected curing conditions and binder [3,25]. In general, it is accepted that the more stable the hydrate is, the higher its dehydration temperature. Consequently, as the water vapor pressure increases as a function of the temperature (Eq. 1), dense refractories are prone to undergoing spalling when pressurization takes place due to hydrate decomposition. This is the reason why the design of refractory products with suitable permeable levels is a key aspect for safe and fast drying as their resulting green mechanical strength may not be high enough to withstand higher thermo-mechanical stress.

Additionally, the ceramic lining for industrial furnaces is usually dried following a one side heating procedure, which gives rise to a hot face (where steam is generated) and a cold region that should release the gas phase. The temperature control is the main parameter that regulates the progress of the drying process and it is often based on the exhaust gas temperature, which might be equivalent to the lining's surface temperature. As reported by Potschke [30], some refractory producers recommend that, after curing, the consolidated ceramic structure should be heated up at 15°C/h with a holding time of 1 h/0.01 m thickness at 150°C, 350°C and 600°C. The selection of these soaking temperatures is related to the water ebullition stage and decomposition of the hydrated phases derived from the binders. Due to the important role of such components in refractory formulations, the following section will highlight the characteristics and main transformations associated with binder agents.

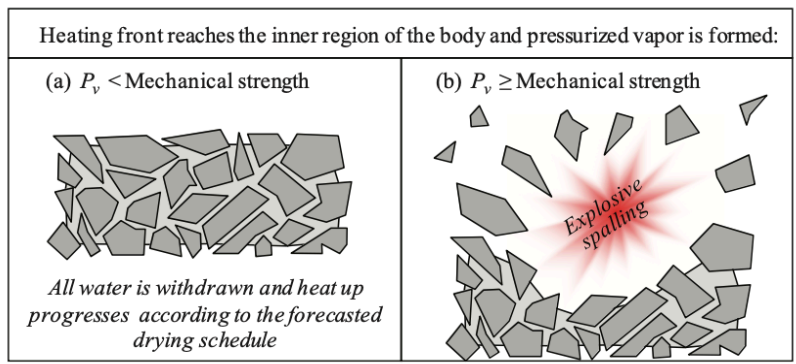

Figure 6: Effect of vapor pressurization and its correlation with the castable's mechanical strength (adapted from [1,15]).

## 2.2. Phase transformations of some binding systems

### 2.2.1. Hydraulic binders

#### 2.2.1.1. Calcium aluminate cement (CAC)

Calcium aluminate cement (CAC) is the most representative hydraulic agent and its most appealing features in industrial applications are fast setting and ability to provide high green mechanical strength (within the curing time of 6 to 24h) and resistance to chemically aggressive environments [3,31–33]. This binder can be produced by reaction of lime ($CaCO_3$) and alumina ($Al_2O_3$) either by a sintering/clinker process at 1315-1425°C or fusion at 1450-1550°C and both lime/alumina ratio and temperature will define the amount and type of the resulting calcium aluminate phases (i.e., mainly $C_{12}A_7$, CA and $CA_2$, where C = CaO and A = $Al_2O_3$) [3,34,35]. Depending on the amount and proportions of these formed compounds, the overall reactivity of the binder will be different. For instance, cements rich in CaO commonly induce faster settings and better green mechanical strength, whereas increased alumina content leads to products with higher refractoriness [3,32,34,36].

Although CAC hydration has been widely studied over the years [33,35–37], this process is still not fully understood due to its complexity. It is accepted that three main steps can be identified during the cement reaction with water: dissolution, nucleation and precipitation. Initially, hydroxylation of the CAC particles' surface takes place leading to its dissolution and release of $Ca^{2+}$ and $Al(OH)_4^-$ ions into the liquid medium (Eq. 2). After that, the latter ion can also be dissociated into $Al^{3+}$ and $OH^-$ species (Eq. 3) [32,34].

$$CaAl_2O_4 + 4H_2O \rightarrow Ca^{2+} + 2Al(OH)_4^- \qquad (2)$$

$$Al(OH)_4^- \rightarrow Al^{3+} + 4OH^- \qquad (3)$$

Considering the progress of the ion formation due to the cement dissolution, the suspension should reach its saturation point, giving rise to a suitable condition to allow the beginning of the nuclei generation of the hydrated phases. This stage, which is called "induction period", is the longer one and it remains until the first crystalline hydrate nuclei is formed [32,36,38]. After that, precipitation is quickly carried out as a result of crystal growth.

Different hydrates can be generated during the CAC reaction with water depending on the cement composition, the selected curing temperature and time, as well as the liquid content added to the mixture. Fig. 7 points out the main transformations expected to take place during the hydration of phases CA and $CA_2$, indicating that the formation of the less soluble and more stable hydrates (i.e., $C_3AH_6$ and $AH_3$) can be favored when using higher temperatures during curing.

$CAH_{10}$, $C_2AH_8$, $C_4AH_x$ (where x = 7, 11, 13 or 19) and $AH_{3(gel)}$ are metastable components that can be dissolved and precipitated into stable $C_3AH_6$ and crystalline $AH_3$ when increasing time and temperature and depending on the environment humidity [34,38]. These transformations (conversion of the less stable hydrates) are usually followed by substantial porosity generation due to the higher density of the most stable phases [3,38]. Hence, one should be aware that the curing conditions of the CAC-bonded castables not only influence the resulting green mechanical strength of these products, but also affect their dewatering behavior as a consequence of the microstructural changes and release of the combined water, derived from the formed hydrates, during heating treatments [39]. Table 1 shows the decomposition temperatures of some of the main calcium aluminate cement hydrates. In general, the dry out schedules (i.e., selection of the heating rates and holding time) of CAC-containing compositions are designed taking into consideration the temperature ranges presented in Table 1, as the uncontrolled release of the chemically bonded water of these hydrates might increase the explosive spalling likelihood of dense refractories.

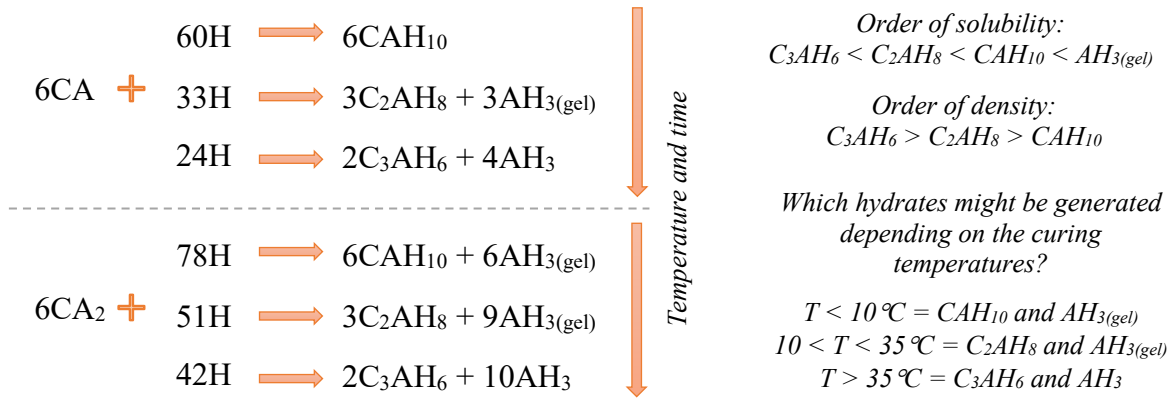

Figure 7: General equations representing the hydrate generation reactions for the main calcium aluminate cement phases. C = CaO, A = $Al_2O_3$ and H = $H_2O$ (adapted from [36]).

Table 1: Decomposition temperatures of some of the calcium aluminate cement hydrated phases [39].

| Hydrated phases | Decomposition temperature (°C) |
|---|---|
| $CAH_{10}$ | 120 |
| $C_2AH_8$ | 170-195 |
| $C_3AH_6$ | 240-370 |
| $AH_3$ (gibbsite) | 210-300 |
| $AH_3$ gel | 100 |

2.2.1.2. Hydratable alumina (HA)

Hydratable alumina is a hydraulic binder produced by the flash calcination of gibbsite, which results in a product comprised by a mixture of rho-$Al_2O_3$ (> 88 wt.%) and other alumina-based transition phases. This powder (usually showing a specific surface area ranging from 100 to 200 m$^2$/g) hydrates in the presence of water, as indicated in Eq. 4, inducing the generation of a thick layer of alumina gel on the particles' surface. This amorphous phase may represent up to 60% of the final hydrates (depending on temperature and pH) and it is, in a further moment, converted into boehmite [$Al_2O_3.(1-2)H_2O$] and bayerite ($Al_2O.3H_2O$) during the castables' curing step. Therefore, the green mechanical strength of the HA-bonded refractories is attained by the interlocking of bayerite crystals and the gel phase that will be scattered in the microstructure and may fill in the available pores [40,41].

$$2\rho - Al_2O_3 + nH_2O \rightarrow 2Al(OH)_3 + Al_2O_3.(n-2)H_2O \qquad (4)$$

In general, hydratable alumina binders show some drawbacks when compared to CAC, such as: *(i)* the need for longer mixing times, *(ii)* higher water demand during the castable processing, which may lead to the risk of adding more liquid than that recommended to ensure proper flowability, and *(iii)* the low permeability of the molded monolithic pieces due to the generation of crystalline and gelatinous alumina phases that clog the available pores and reduce the number of permeable paths contained in the formed microstructure. Thus, longer drying schedules and paying special attention to the 80-120°C and 210-300°C temperature ranges must be taken into consideration when heating HA-bonded refractories.

For example, Fig. 8 shows the drying rate curves (dW/dt or DTG) of CAC and HA-bonded castables (containing 2 wt.% and 4.5 wt.% of binder and water, respectively) when subjected to a heating rate of 10°C/min. A first peak related to evaporation was observed, followed by a second one that corresponds to both ebullition and hydrate decomposition. The intensity of the evaporation peak for the HA sample (ii) was lower than the correspondent one for CAC-based composition (i) meaning that a greater volume of free water was retained in the microstructure. This indicates that the former has a higher potential of pressurization during ebullition [42]. Moisture-free samples (MF, previously dried in silica gel containing environment prior to the thermal treatment) for the same compositions were also analyzed and presented in Fig. 8.

According to profiles (iii) and (iv), a considerable amount of the hydrated phases decomposed in the 100 to 250-350°C interval, but CAC-MF showed three decomposition peaks, whereas HA-MF presented a more continuous range of water loss. The drying behavior of hydratable alumina containing samples is mainly influenced by the resultant permeability, as this property inhibits the water from withdrawal during the evaporation stage and may lead to a harmful vapor build up during ebullition [42].

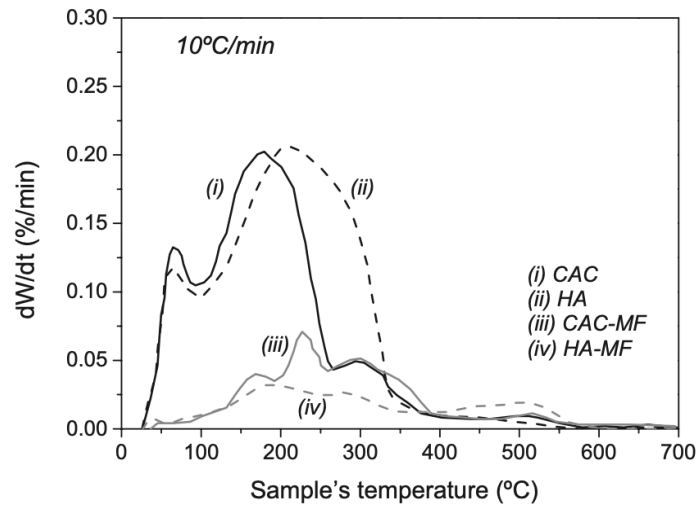

Figure 8: Drying behavior of green castables bonded with 2 wt.% of CAC or HA submitted to continuous heating rate of 10°C/min. MF = moisture-free samples [42].

Despite the importance and the extensive use of CAC and HA, ageing (due to interactions between these binders or castable dry-mix and atmosphere) has been pointed out as an intrinsic difficulty for the monolithic industry, resulting in negative effects to the overall performance of these products [43–45]. Consequently, using binders less susceptible to this phenomenon and to the variation of the processing parameters (water/binder content, humidity, time and temperature) than CAC is a subject under constant investigation [44,46,47].

2.2.1.3. Magnesium oxide

The development of MgO-bonded castables has been continuously explored [20,23,56,48–55] due to the challenges and concerns for producing high quality and crack-free compositions. In such products, the bonding effect is derived from *in situ* Mg(OH)$_2$ (brucite) formation (Eq. 5) in the interstices and voids among the coarse and fine particles of the refractory [16,57]. However, this phase transformation must be followed carefully as it results in a 2.5-fold expansion due to the density mismatch between magnesia ($\rho$ = 3.5 g/cm$^3$) and brucite ($\rho$ = 2.4 g/cm$^3$), which may induce the formation of cracks in the consolidated refractory pieces [25,48].

$$MgO + H_2O_{(l,v)} \rightarrow Mg(OH)_2 \quad (5)$$

Among the various parameters that can affect MgO hydration during the castables' processing steps, the magnesia source (i.e., caustic, dead-burnt or electrofused ones, their granulometry, specific surface area, purity, CaO/SiO$_2$ ratio, and others) and its interaction with other compounds are pointed out as the two most important aspects. In general, fused magnesia is not a suitable binding agent because it is comprised by large crystals with limited reactivity (Fig. 9). On the other hand, the high reactivity of caustic MgO may generate an excessive amount of Mg(OH)$_2$ during water mixing and curing processes of castables. Thus, in addition to selecting magnesium oxides with proper physical-chemical features, one may focus on inducing a better control of brucite formation during the refractory processing steps [58–60].

Aiming to take advantage of the MgO hydration and optimizing the bonding potential of this material, some approaches are suggested in the literature, such as: *(i)* inducing faster Mg(OH)$_2$ formation before the composition setting time when the molded castable still has enough room and some freedom to accommodate stresses [16,50], *(ii)* changing the morphology of the hydrated phase to better fit the crystal growth in the designed microstructure [58–60], and *(iii)* using hydrating agents (i.e., carboxylic acids, and others) to activate a higher number of sites and increase the Mg(OH)$_2$ nucleation rate on the MgO surface, which might limit the crystal growth [54,55].

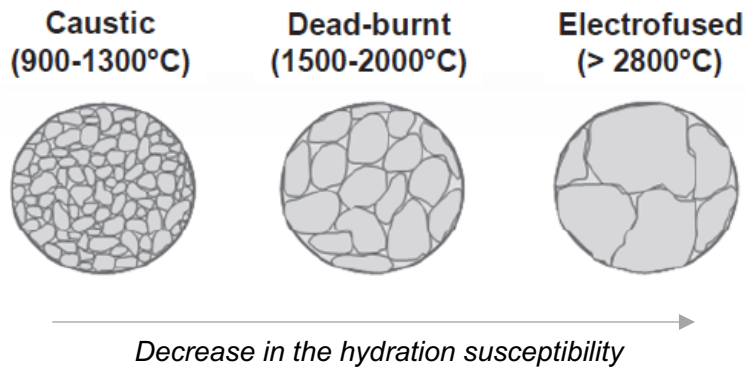

Figure 9: Temperatures used during the thermal treatment of MgCO$_3$ for the production of different MgO sources and their effects on the resulting primary crystal size (adapted from [25,61,62]).

Knowing that MgO can readily react with water in its liquid or vapor form, Mg(OH)$_2$ generation can take place at the initial stages of the refractories' drying process, which represents an additional risk to the integrity of compositions containing this binder. Besides that, brucite decomposition takes place at relatively high temperatures (380-420°C) [62–64], which coupled to the reduced permeability of the resulting microstructure, may result in high steam pressure levels (as predicted by Antoine's equation, Eq. 1) and, consequently, the explosion of the castables even under usual heating schedule profiles. If MgO is exposed to steam for long periods of time, it will also be more prone to rehydration, increasing the likelihood of the cracking formation in the castable structure. Therefore, water withdrawal in magnesia-bonded castables must be carried out as fast as possible because due to the low porosity and permeability of the designed microstructures, the steam release associated with brucite decomposition might be hindered [26,55].

Fig. 10 illustrates the drying rate profile of high-alumina MgO-bonded castables when the green samples were submitted to heating rates of 5 and 20°C/min [20]. As observed, the reference composition containing 94 wt.% of alumina and 6 wt.% of MgO did not withstand the stresses derived from the ebullition stage and exploded in both tested conditions at temperatures close to 150°C (before brucite decomposition). As the evaluated materials were only cured at 30°C/24h before testing, a large amount of water was still present in the microstructure during the heating step. When adding an organic salt (aluminum salt of 2-hydroxypropanoic acid) and a permeability enhancing compound (Refpac® MiPore 20) as drying agents to this castable, the microstructural changes resulting from the action of these raw materials (i.e., gel-like phase decomposition and interlamellar water withdrawal from hydrotalcite-like phases [20]) allowed the release of a higher amount of water in the 80-123°C range (the most intense peak shown in Fig. 10b). The characteristic brucite decomposition peak (402 or 403°C) could also be detected for the castable containing the organic salt.

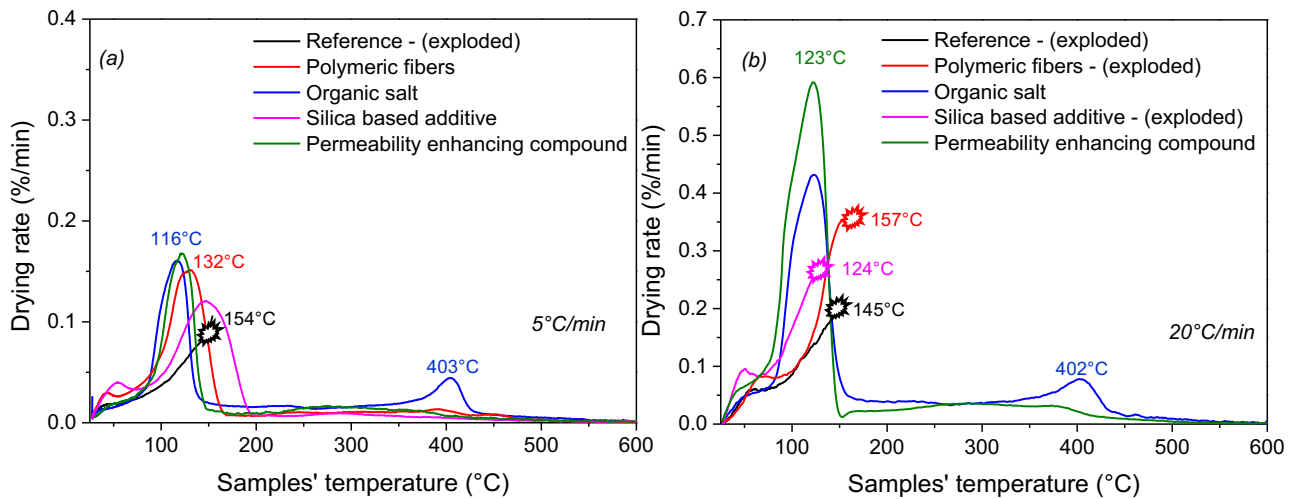

Figure 10: Drying rate of MgO-bonded castables for thermal treatments carried out with a heating rate of (a) 5°C/min and (b) 20°C/min. All samples were cured at 30°C for 1 day before the measurements (adapted from [20]).

On the other hand, the incorporation of polymeric fibers (polyethylene and with ~3 mm diameter and length of 6 mm) or a silica-based additive (SioxX®-Mag) into the magnesia-bonded composition was only effective to prevent the explosion of samples when the heating treatment was carried out with 5°C/min rate (Fig. 10a). Hence, in order to prevent earlier cracking of the cured samples and induce faster steam withdrawal during the first heating treatments of the molded pieces, it is important to find alternatives to control MgO hydration and improve the permeability of refractory dense structures [20].

2.2.2. Colloidal silica

Many studies have been carried out in recent years regarding using nano-powders and colloidal suspensions in refractory compositions, aiming to improve bonding and densification at lower temperatures and to optimize their drying step, as the setting mechanism of these nano additives usually induces the generation of porous and highly permeable structures [24,46,47,65,66].

Unlike the hydraulic binders, the consolidation mechanism of nano-binders does not result in hydrated phases and the total liquid content required for suitable mixing of the castables is defined by taking into account the water amount available in the colloidal suspensions [25,46]. Colloidal silica (CS) can induce the setting of high solid loading mixture into pieces by the formation of a gel (gelling mechanism) derived from the siloxane bonds generated among the $SiO_2$ particles (Eq. 6) [47].

$$\equiv SiOH + HOSi \equiv \rightarrow \equiv SiOSi \equiv + H_2O \tag{6}$$

On the other hand, the flocculation of CS-containing refractories can be favored when an additive bridges the particles, giving rise to close-packing clumps. Dead-burnt MgO has been commonly used as the main setting agent in castables containing colloidal silica [46,47,67] as the hydration of magnesia withdraws hydrogen ions from Si-OH groups, inducing a higher siloxane bond formation. Conversely, when salts (such as $CaCl_2$, $MgCl_2$, $MgSO_4$, etc.) are added to CS-based suspensions, the ions derived from these additives act as coagulating agents, reducing the overall net repulsion effect and increasing the gelling rate [47,67,68]. Both mechanisms (flocculation and gelling) are influenced by the presence of electrolytes, pH, particle size and concentration of silica, and temperature [46,68].

Castables bonded with colloidal silica are usually less sensitive than CAC, HA and MgO to curing conditions (time and temperature) due to its consolidation mechanisms. Nevertheless, the usual setting times are still long, and the as-cast green mechanical strength of CS-containing refractories are lower than those attained for CAC-based systems. For the sake of comparison, Fig. 11 presents the cold crushing strength of high-alumina castables containing 5 wt.% of hydratable alumina, calcium aluminate cement (binder content: LCC-FS = 5 wt.%, ULCC-FS = 2 wt.% and LCC-A = 6 wt.%, and with 5 wt.% or without fumed silica - FS) or colloidal silica (8 wt.% and concentration = 40%) after keeping the prepared samples at temperatures in the range of 10 to 30°C for 24h.

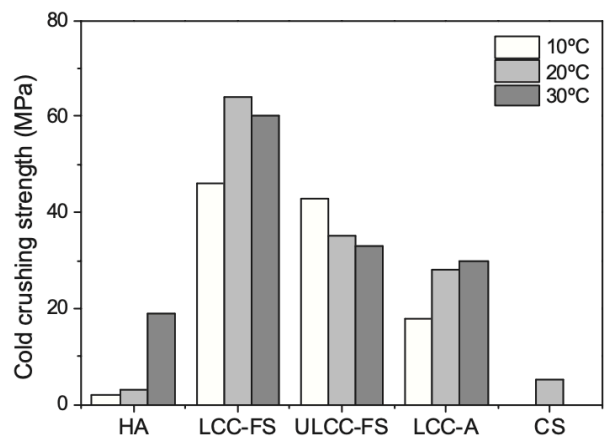

Figure 11: Cold crushing strength after 24h at different temperatures (10, 20 or 30°C) of high-alumina castables containing hydratable alumina (HA), calcium aluminate cement [LCC = low cement containing and ULCC = ultra-low cement containing (FS) or without (A) fumed silica] or colloidal silica (CS) as a binder source [69].

According to Parr et al. [69], the nano-bonded composition pointed out in Fig. 11 presented very low mechanical strength even when the samples were exposed to a longer curing time (> 48h) at 20°C. In this case, the crushing strength values were close to 5 MPa, which was almost 75% lower than the low cement containing composition (LCC-A). Other studies highlighted that combining CS with hydraulic binders may also be a suitable alternative to overcoming this issue and optimizing the green mechanical properties of nano-bonded refractories [24,70].

Colloidal silica-bonded refractories usually present a high permeable and porous structure [69,71,72], which results in the previously mentioned decrease in the overall green mechanical strength of the cast pieces. On the other hand, these features can be useful in terms of reducing the risks of explosive spalling during the first heat-up of these castables. Fig. 12 shows the drying rate of castables bonded with colloidal silica, hydratable alumina or the blend of both binders (the mixture of CS and CAC is not indicated, as this system tends to coagulate due to different pH ranges of these binders when in contact with water). As the CS binder does not give rise to hydrated phases, most of the water retained in the pores was released during evaporation and at the beginning of the ebullition steps (< 150°C), inhibiting the likelihood of the systems' pressurization.

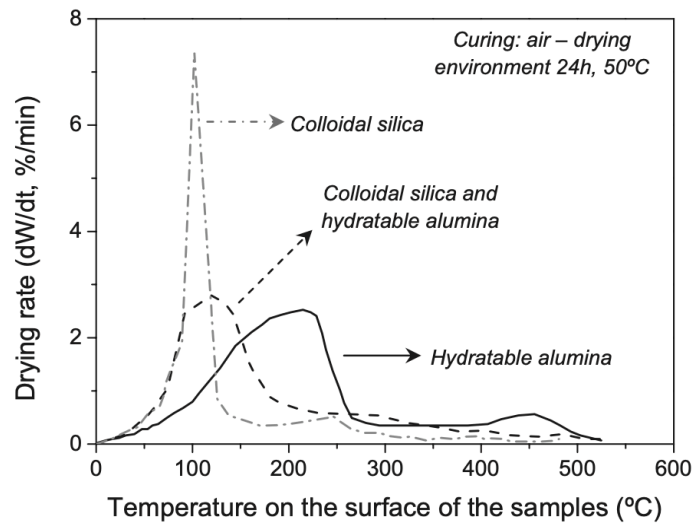

Figure 12: Drying rate of green samples of castables bonded with colloidal silica, hydratable alumina or the blend of both binders [72].

As discussed in Section 2.2.1.2., HA-bonded compositions require the use of lower rates for drying due to the lower permeability and the presence of phases that displaces the water withdrawal to higher temperatures (resulting in a longer time to fully release the free and combined water). Thus, the blend of CS and HA may be interesting because it can lead to shorter drying time when compared to the HA-bonded one and also help increase the green mechanical strength of the obtained refractories [72]. Nevertheless, the search for effective gelling agents for silica sol-bonded castables is still an interesting and important subject to reduce the setting time and improve the green mechanical strength of such systems without spoiling their drying speed.

2.2.3. Chemical binders (phosphates)

Phosphate bonds are of particular interest in the refractory field due to their good adhesion and a reduced risk of cracking under high heating rates after curing. They may be formed by the addition or *in situ* generation of phosphates in the ceramic structure where phosphoric acid ($H_3PO_4$) or monoaluminum phosphate (MAP or $Al(H_2PO_4)_3$] are pointed out as their main sources. Nevertheless, little information regarding the setting mechanisms and chemical reactions of these materials during placing and drying of monolithics is presented in the literature [31,73,74].

Based on the available data, phosphoric acid may react with alumina above 127°C or with aluminum hydroxide at room temperature, giving rise to MAP (Eq. 7 and 8). After some time, $Al(H_2PO_4)_3$ decomposition is usually observed and followed by precipitation of $AlPO_4 \cdot xH_2O$ or $[Al(PO_3)_3]_x$ that induces the setting of the processed monolithic products [25,31].

$$6H_3PO_4 + Al_2O_3 \rightarrow 2Al(H_2PO_4)_3 + 3H_2O \qquad (7)$$

$$3H_3PO_4 + Al(OH)_3 \rightarrow Al(H_2PO_4)_3 + 3H_2O \qquad (8)$$

Despite the high number of available aluminum-based phosphate compounds (more than 50), MAP is still the most applied chemical binder in the refractories due to its solubility in water, high bonding strength and reaction with basic and amphoteric raw materials at low temperatures [75–77]. According to Luz et al. [78], different routes can induce *in situ* MAP generation in castable compositions, such as: (i) phosphoric acid incorporation into dry-mixes containing $Al_2O_3$ or $Al(OH)_3$, or (ii) mixing alumina and/or aluminum hydroxide particles with the $H_3PO_4$ solution before the castables' processing step, or (iii) favoring the $Al(OH)_3$ generation during the castable's mixing stage by adding hydratable alumina to the formulation, which may enable an effective interaction among $H_3PO_4$-hydroxide-alumina during the preparation of such refractories.

It is important to highlight that high-alumina MAP-bonded refractories do not usually present fast hardening at room temperature and using setting agents in such a system should be considered to control their workability and stiffening behavior. For instance, MgO, CaO, calcium aluminates and others can induce an acid-base exothermic reactions when in contact with $H_3PO_4$ and MAP, resulting in amorphous/crystalline phosphates that affect the refractory hardening and other properties [31,75]. In the case of magnesia-containing phosphate bonded castables, the bonding effect is derived from the interaction of this oxide with phosphoric acid (Eq. 9), which gives rise to a metastable and soluble phase [$Mg(H_2PO_4)_2 \cdot H_2O$] during the initial processing stages of such materials. After that, this hydrate or MAP may also react with MgO and $H_2O$ available in the composition (Eq. 10-11), resulting in the formation of $MgHPO_4 \cdot 3H_2O$ (newberyite) and $AlPO_4 \cdot nH_2O$ [31,74].

$$MgO + 2H_3PO_4 \rightarrow Mg(H_2PO_4)_2 \cdot H_2O \qquad (9)$$

$$Mg(H_2PO_4)_2 \cdot H_2O + MgO + H_2O \rightarrow 2MgHPO_4 \cdot 3H_2O \qquad (10)$$

$$2MgO + Al(H_2PO_4)_3 + (n+1)H_2O \rightarrow 2MgHPO_4 \cdot 3H_2O + AlPO_4 \cdot nH_2O \qquad (11)$$

Consequently, the setting time and workability of such refractories depend on the MgO features (crystallinity degree, surface area, grain size, its added content, etc.), as well as the concentration of the phosphate-based solutions.

Despite the hydrates' generation in the microstructure of the chemically-bonded refractories, it is accepted that these products present suitable spalling resistance even when they are exposed to high heating rates. This might be attributed to the early decomposition of the formed phases, as well as the consolidated porous structure resulting from the intense interaction between the acid solution and basic oxides. As shown in Fig. 13, the drying rate profile of the matrix fraction of high-alumina phosphoric acid-bonded castables containing dead-burnt magnesia attested the decomposition of newberyite and $AlH_3(PO_4) \cdot H_2O$ in the 80-200°C temperature range. Moreover, these castables presented overall apparent porosity close to 20% after drying at 110°C [78].

Therefore, a key aspect to produce phosphate-bonded castables with enhanced performance consists of selecting raw materials with proper reactivity and solubility because the phase transformations that take place during the curing stage must evolve under a compatible rate, favoring the organization/accommodation of the resultant products to form a cohesive microstructure.

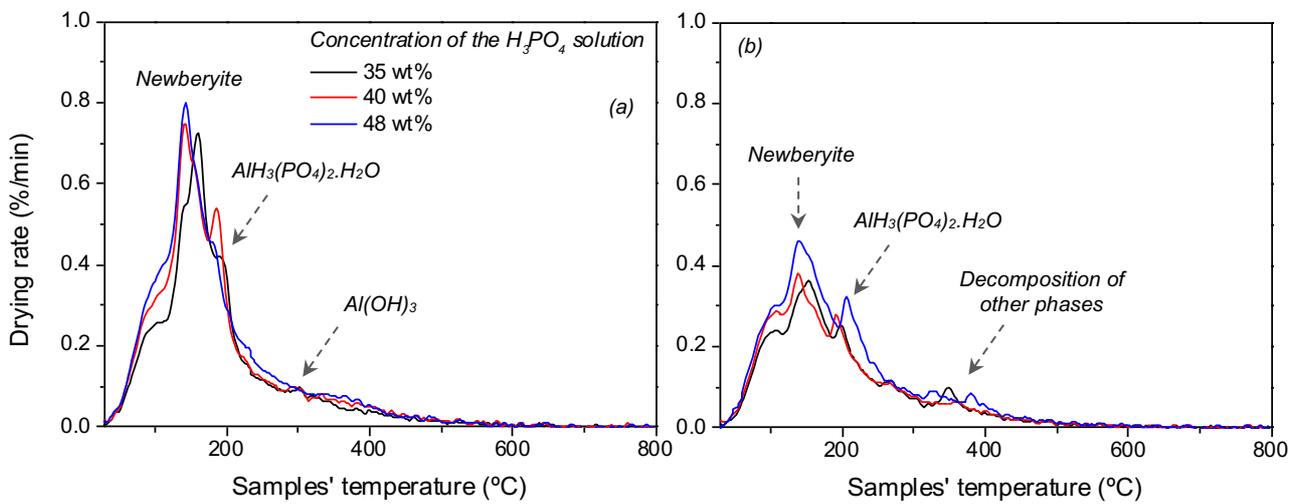

Figure 13: Drying rate of the matrix composition of high-alumina castables containing phosphoric acid and (a) hydratable alumina (added to the dry-mix) or (b) aluminum hydroxide (previously added to the acid solution). Samples prepared with $H_3PO_4$ solution with different concentrations (35-48 wt.%) and dried at 110°C for 24h were analyzed in these tests [78].

Aiming to point out the most applied experimental techniques used to evaluate the drying behavior of refractory castables, the following sections will present and discuss the features and main data that can be obtained in such measurements.

*2.3. Experimental techniques used to assess the water removal of dense ceramics*

2.3.1. Thermogravimetric and explosion tests

A classical way to study the kinetics of the binders' dehydration and the drying behavior of ceramic materials consists of carrying out thermogravimetric analysis (TGA) and differential thermal analysis (DTA). TGA allows the evaluation of the main dewatering steps of the specimens by calculating the normalized $W$ parameter, which measures the cumulative fraction of water expelled during the heat up divided by the total $H_2O$ content initially contained in the body (Eq. 12). The derivative (DTG or $dW/dt$) is also calculated to obtain the mass loss/drying rate at each time $t_i$ (Eq. 13).

$$W(\%) = 100 \times \left(\frac{M_0 - M}{M_0 - M_f}\right) \qquad (12)$$

$$\left(\frac{dW}{dt}\right) = \frac{W_{i+1} - W_{i-1}}{t_{i+1} - t_{i-1}} \qquad (13)$$

where $M_0$ is the initial mass, $M$ is the instantaneous mass recorded at a certain time during the heating stage and $M_f$ is the final (dry) mass of the sample.

As shown in Fig. 14, the different stages (the first three presented peaks are associated with free-water release, $AH_3$ and $C_3AH_6$ decomposition, respectively) of the dehydration process of a hydrated cement-based matrix can be inferred with thermogravimetric measurements. However, the commercially available TGA equipment usually carries out such measurements using small solid specimens or, more frequently powdered ones with mass values in the range of 10 mg to 2 g [37,79]. Knowing that most of the castable compositions comprise coarse grains that are even larger than the small crucibles used in such devices and that the samples' sizes will play an important role in the development of thermal gradients and pressure build-up in the resultant microstructure of such materials, various laboratories had to adapt and create their own setup (sometimes called "macro"-TGA [80]) to carry out more representative and reliable tests for the characterization of refractory products.

These tests are usually conducted in an electric furnace controlled by a proportional-integral-derivative (PID) system and the mass loss of the samples are recorded by a laboratory scale [5,29]. Cylindrical or cubic samples with sizes varying from 30 mm x 30 mm up to 100 mm x 100 mm are suspended at the center of the furnace. Mass loss and temperature of the specimens may be simultaneously measured using thermocouples that are embedded close to the surface or at the center of the refractory specimens during their casting. As water vaporization is an endothermic process, monitoring the samples' temperature can be a feasible way of following the drying front progress of the refractories. Hence, the "macro"-TGA tests are carried out in the 20-600°C range and with heating rates varying from 2-20°C/min [5,29,80,81]. As an example, Fig. 15 shows the mass loss (TG) and DTG profiles obtained for cured cylindrical high-alumina castable samples (50 mm x 50 mm) when a heating rate of 20°C/min is applied during their first thermal treatment. As observed, the explosion and temperature at which this event took place could be detected during the evaluation of such refractories. Additionally, the temperature range was identified where a higher water withdrawal took place during the specimens' drying, which can be useful to compare compositions containing different raw materials [82].

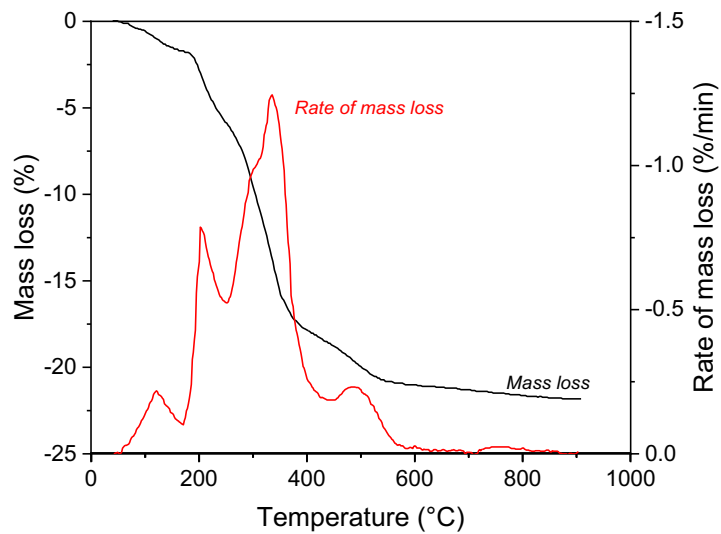

Figure 14: Mass loss and rate of mass loss of a cement-based matrix (water/cement ratio = 0.3) obtained by TGA [79].

Whereas small samples can be fully damaged due to spalling, it was observed in practice that the explosion usually takes place at depths smaller than 100 mm from the castable's surface for thicker and bigger bodies [31]. Some authors pointed out that the faster the heating rate, the earlier and closer to the surface the location of the maximum pore pressure will be [83]. Besides that, when the size of the refractory specimen is larger, the vapor has a longer path to migrate through the microstructure before reaching the surface. Thus, the overall resistance to fluid flow of the packed structure rises with the body's size, which is a further challenge for the dewatering of thick blocks.

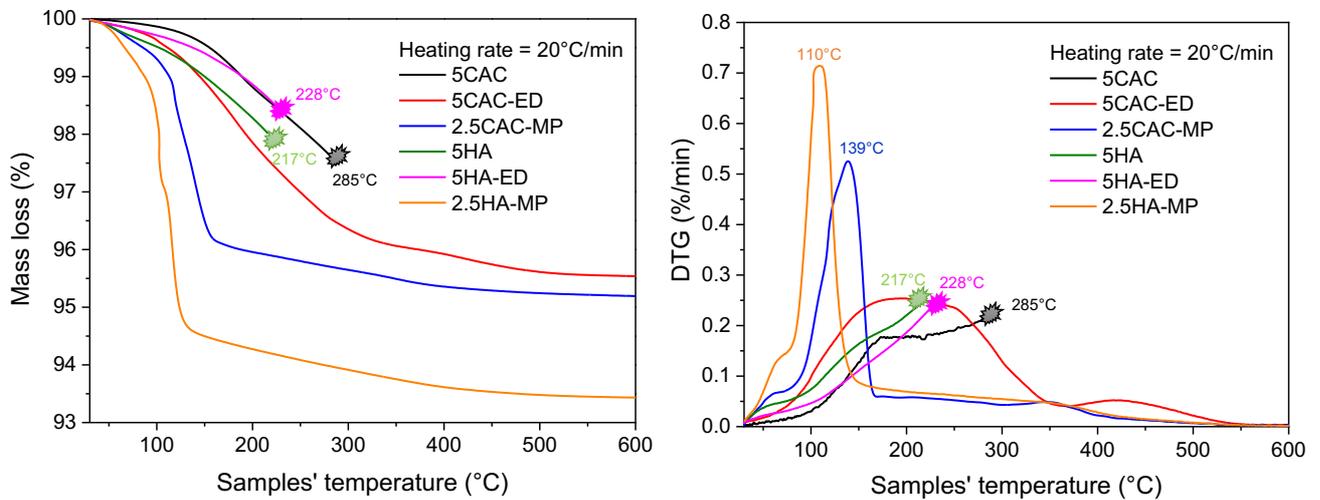

Figure 15: Mass loss profiles and derivative of the mass loss (DTG) as a function of the samples' temperature. The evaluated castables were bonded with calcium aluminate cement (CAC) or hydratable alumina (HA). Results collected for reference compositions (without a drying additive = 5CAC and 5HA), as well as the ones containing polymeric fibers (ED) or a permeability enhancing active compound (MP), are shown [82].

Thus, laboratory and industrial-scale explosion resistance measurements have been proposed using 100 mm x 100 mm x 100 mm cubes and larger blocks [up to pieces with 600 mm x 600 mm x 350 mm (~400kg)], following the Chinese Standard YB/T4117-2003 [80,84,85]. The applied procedure consists of inserting a specimen inside a furnace held at the desired temperature and then observing if it has been damaged or not after a given period in the furnace. This test assesses the ability of the refractory to release water swiftly and to adapt the microstructural changes due to high thermal stresses under such conditions. Despite the fact that it does not provide any information about the relationships between intrinsic parameters of the castable (i.e., water content, permeability, etc.) and external conditions (curing temperature/time, furnace temperature, specimen size, etc.) to aid the understanding of spalling mechanisms, such measurements highlight the temperature dependent sensitivity for the refractory and can be used for a comparison when changing formulation parameters, as indicated in Table 2 [80,86].

Table 2: Results of explosion resistance tests (samples were directly inserted into the furnace at the selected temperature). Al stands for aluminum powder [86].

| Material | | Anti-explosion additives | | |
|---|---|---|---|---|
| | | Reference (additive-free) | Organic fiber | Al powder |
| Water demand (%) | | 7.5 | 7.5 | 7.5 |
| Curing temperature (°C) | | 10 | 10 | 10 |
| Furnace temperature (°C) | 650 | O O O | O O O | O O O |
| | 700 | × × × | O O O | O O O |
| | 750 | × × × | O O × | O O O |
| | 800 | × × × | × × × | O O O |

Specimen size: cylinders of 80 mm x 80 mm. Spalling: O (not occurred), × (exploded).

In order to detect the temperature at which water withdrawal or the samples' explosion took place, it is very common also to carry out such tests with thermocouples embedded at the surface and/or at the center of the castables. Fig. 16 presents some details of the information that can be collected by these measurements and highlights the fact that to ensure the safety of the refractory lining of the furnace, the evaluated samples (100 mm x 100 mm x 100 mm cubes) were placed in a metallic box able to resist an explosion. The collected temperature profiles of a CAC-bonded castable points out that the analyzed specimen survived the heating conditions (20°C/min) and the typical stages [evaporation, hydrate recombination (conversion) and dehydration] described for dry-out of low-cement castables could be identified [80].

Nevertheless, a more dynamic approach, close to the actual heating condition applied in industrial practices, would consist of exposing just one side of the refractory material (the hot face) to the temperature [6,80,86]. Takazawa et al. [86] called this procedure "panel method", where the castables' explosion resistance can be analyzed by placing them at different positions (simulating the ceiling, sidewall and floor) during testing. Moreover, the samples' sizes can be as large as 250 mm x 250 mm x 200 mm and 1100 mm x 500 mm x 200 mm. As shown in Fig. 17a, thermocouples may be placed at different positions to record the temperature: *(i)* of the furnace ($T_1$) as this device was used to control the heating rate applied during the measurements, *(ii)* of the dense refractory at the panel surface ($T_2$), *(iii)* at the interface between dense and insulating refractories ($T_3$), and *(iv)* of the steel shell ($T_4$).

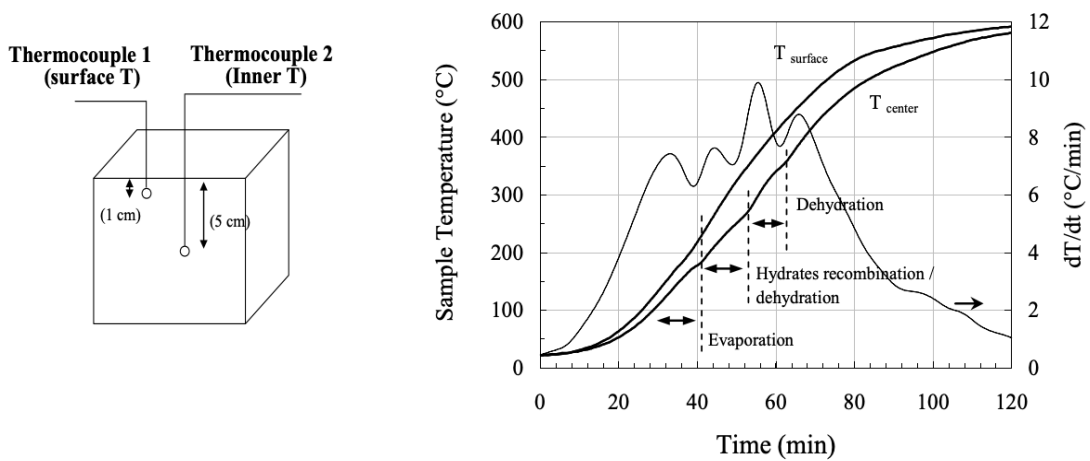

Figure 16: Details of the refractory sample and the explosion test proposed by Auvray et al. [80], pointing out the typical temperature profiles obtained at the surface and center of a CAC-bonded castable, when a 20°C/min heating rate was applied.

Takazawa et al. [86] also proposed some adjustments in the tested specimens to prevent their spalling during the panel tests. The following modifications were analyzed:
- Case 1 = Bamboo sticks were inserted at 280 mm deep in the molded panels during their casting (Fig. 17a), aiming to produce 4 venting holes (after burning out the sticks) separated at intervals of 200 mm.

- Case 2 = Steel rods with 10 mm diameter were used instead of the chopsticks. The venting holes were formed by pulling out these rods after demolding the samples.
- Case 3 = The same procedure described in case 1 was applied at the steel shell (located behind the lightweight and dense refractories), resulting in the presence of four holes of 10 mm diameter at a distance of 200 mm from each other.
- Case 4 = a panel comprised by only a monolayer of the high-alumina based castables was tested. The same procedure described in case 1 was used.

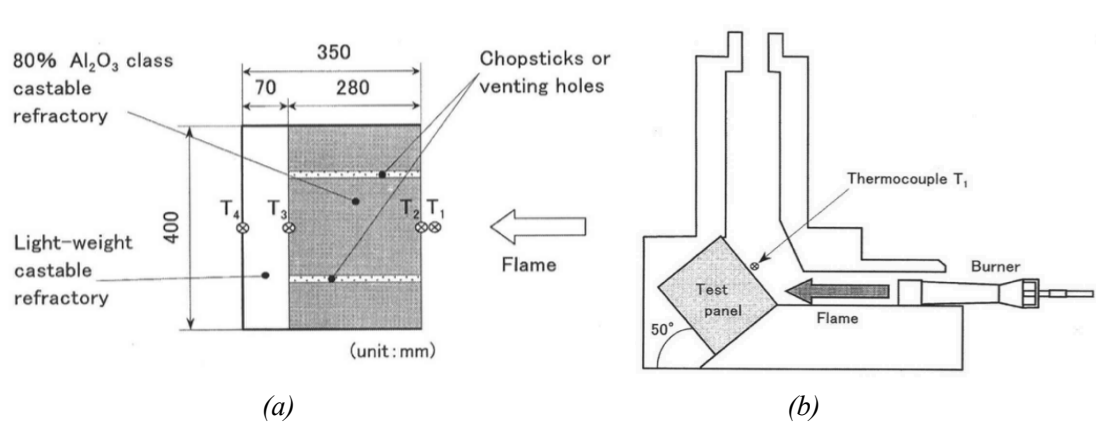

Figure 17: (a) Details of the refractory samples prepared for the "panel test" and (b) sketch of the setup used for those explosive spalling tests [86].

Table 3 summarizes the results of these explosive spalling tests for three heating conditions (holding time of 0, 2 and 4 hours) and four installation cases. The panels were heated up to 1000°C using a gas burner (Fig. 17b) and a heating rate of 200°C/h was assumed. The explosion of the samples (case 1, 2 and 4) was observed always after the burner was extinguished. The presence of the holes in the steel shell (case 3) prevented the materials explosion and affected the measured temperature at the refractories' interface ($T_3$ increased from 70-80°C up to 110°C after the panels' surface reached 1000°C). This $T_3$ increase suggested that fast water vaporization through the holes in the shell led to an improved steam diffusion and thermal flux in the evaluated samples. Consequently, the pressure build-up could be better controlled in case 3.

Table 3: Results of the explosive spalling test (panel method) as proposed by Takazawa et al. [86].

| Holding time at 1000°C (h) | Test conditions | | | |
|---|---|---|---|---|
| | Case 1 | Case 2 | Case 3 | Case 4 |
| 0 | × | × | ○ | × |
| 2 | × | × | ○ | ○ |
| 4 | ○ | ○ | ○ | ○ |

Spalling: ○ (not occurred), × (exploded).

Therefore, the development of monitoring tools to assist the dry-out of refractory castables, coupled with an experimental validation may help to successfully design optimized heat-up schedules and safely dry-out refractory castables. Mass loss, temperature, and internal vapor pressure profiles are pointed out as some of the most important parameters required for constructing numerical drying models to predict the likely heating schedule that may result in a lower risk of explosion for the ceramic lining [12,14,80].

2.3.2. Microwave heating

Microwaves are electromagnetic waves with frequencies ranging from 300 MHz to 300 GHz (wavelengths from 1.0 mm to 1.0 m), but most of the available systems operate between 433.92 MHz and 40 GHz. In the ceramic industry, microwaves are mainly used for drying and heating ceramics with the advantage that the electromagnetic waves penetrate into the produced pieces and heat their entire volume [87,88].

Microwave heating is related to the materials' ability to absorb energy when placed in a high-frequency electric field. For instance, dielectric ceramics, comprised by polar molecules with positive and negative poles, should present dipole polarization and conduction when exposed to the high-frequency electric field of the microwave, resulting in instantaneous vibrations of these orderly dispersed molecules (Fig. 18). Consequently, heat is produced in the materials' entire volume due to the molecular friction [88,89].

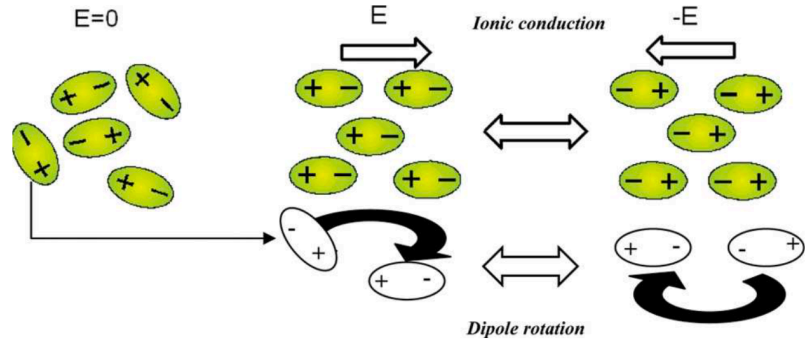

Figure 18: Sketch of the microwave heating mechanisms [88].

The resulting heat or the microwave energy absorbed is described by Lambert's law (Eq. 14) and it depends directly on the wave frequency and dielectric property of the evaluated material [88,89]:

$$Q = \sigma |\vec{E}|^2 = 2\pi f \varepsilon_0 \varepsilon_r' (\tan \delta)|\vec{E}|^2 \qquad (14)$$

where $Q$ is the microwave energy, $\sigma$ is the effective conductivity [S/m], $f$ is the frequency (Hz), $\varepsilon_0$ is the permittivity of free space (8.8514 x 10$^{-12}$ F/m), $\varepsilon_r'$ is the relative dielectric constant, $\tan \delta$ is the dielectric loss tangent coefficient, and $\vec{E}$ is the electric-field intensity (V/m).

The main advantages of microwave heating compared to conventional methods are [87,88,90–92]: *(i)* application of high heating rates and simultaneous heating of the entire volume of the ceramic pieces, resulting in lower energy consumption and shorter processing times; *(ii)* deep penetration of the electromagnetic waves and better homogenization of the temperature in the material to be dried; *(iii)* instantaneous and precise electronic control; and *(iv)* clean heating process that does not generate secondary waste. Ochiai et al. [90] reported experimental results comparing various drying methods applied for the evaluation of vibratable monolithic refractories. Table 4 highlights the better drying efficiency and the lack of damage in the ceramic lining of a ladle furnace when the microwave heating method was used.

Table 4: Comparison of various drying methods applied to the processing of vibratable refractory castables [90].

| Collected data | Gas burner (conventional method) | Hot air (100-200°C) | Microwave (10-15kW/refractory Ton) |
|---|---|---|---|
| Drying time (h) | 15 – 16 | Not dried after 15 h | 5 – 7 |
| Total input energy ($10^3$ kcal/refractory Ton) | 800 – 1000 | 300 – 400 | 150 – 200 |
| Thermal gradient in the lining (°C/cm) | 15 – 20 | 4 – 5 | 1 – 3 |
| Effect of bursting or peeling by drying | Significant damage | Significant damage | None observed |

Recent studies conducted by Czechowski and Majchrowicz [87,91] analyzed the effects of such an alternative drying technique on strength, porosity and phase composition of CAC-bonded castables prepared with various water/cement ratios. The investigations were divided into two parts, where firstly the water vaporization influence on the properties of the cured samples was analyzed by heating them up to 110°C for 24h (conventional treatment) or using microwave power, which was continuously raised from 30 W to 120 W and kept at these maximum values for 10h. A second paper focused on the study of the castables' dehydration behavior when heating them up to 600°C (conventional drying) or applying microwave radiation with increasing power from 100 W to 1300 W at a rate of 200 W/h.

According to these authors [87,91], despite the changes in the water/cement ratio values of the prepared castables, the analyzed properties (modulus of rupture, cold crushing strength, open porosity) and phase composition of the samples subjected to conventional and microwave drying were comparable. When applying the latter, the temperature evolution in the cured samples (up to ~110°C) was very similar (regardless of the tested refractory, Fig. 19a) and an increasing weight loss was observed for the materials containing greater water content (Fig. 19b). On the other hand, the same castables presented a continuous temperature increase up to 200-220°C after ~ 2.5h, but the subsequent raise of power did not result in the samples' heating and their temperature dropped to around 100°C (Fig. 19c). As the temperature was measured on the surface of the specimens with a pyrometer, the measured drop was attributed to the intense water release from the materials' surface, as well as the latent heat associated with water evaporation. A rapid temperature increase in the refractories was observed only above 800 W and their weight loss decreased gradually, as indicated in Fig. 19d. Based on the samples´ behavior, it was stated that microwave heating can be effectively used for drying refractory castables, but in practice the selected power increase has to be adapted depending on the size of precast pieces.

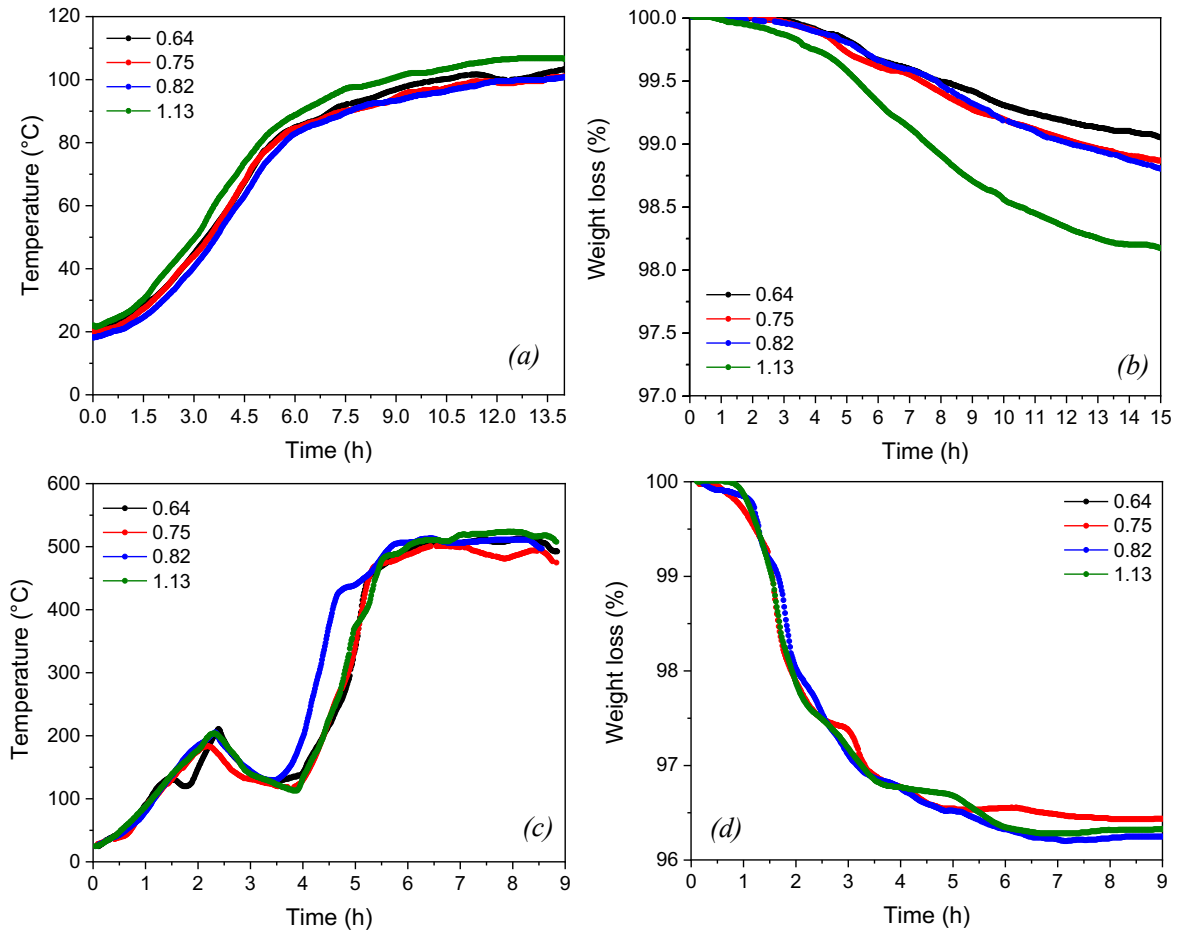

Figure 19: (a,c) Temperature and (b,d) weight loss profiles of LCC-refractory castables when heated with microwave radiation. The caption of the original figures points out the water/cement ratio used during the samples' preparation (adapted from [87,91]).

In order to take advantage of this drying method and reduce the costs for acquiring a microwave power unit, some companies have implemented a combination of two techniques, electromagnetic waves plus hot air supplier, to dry the refractory lining of industrial equipment. Fig. 20 shows a sketch of the processing sequence proposed by Ochiai et al. [90] where the microwaves were transmitted to the ladle linings via waveguides and reflectors during drying.

Based on the presented results, the quality of the refractory castables dried by conventional methods was lower than those obtained by the combination of microwave + hot air heating. Moreover, the authors stated that the working life of the ceramic linings applied at Hirohata Works of Nippon Steel increased around 10-20% after applying this drying procedure [90].

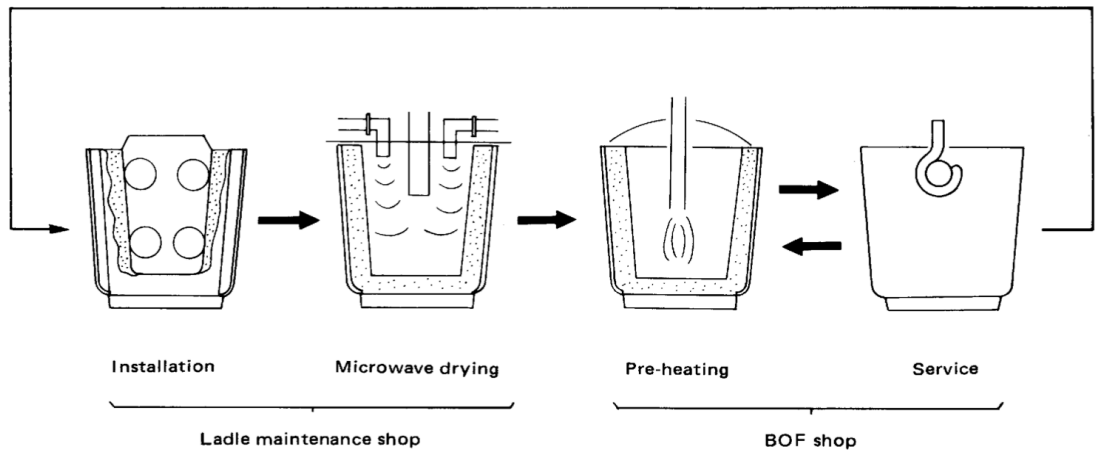

Figure 20: Typical operation cycle of steel ladle considering microwave drying of the ceramic lining [90].

Taira et al. [92] also reported using these combined methods in ladles and RH vessels at ten units of Nippon Steel Corp. Although that paper did not show results from the industrial trials, preliminary tests were carried out for the evaluation of 200 mm x 200 mm x 200 mm or 500 mm x 500 mm x 300 mm alumina-magnesia or alumina-spinel refractory blocks (no information about the binding system was provided), respectively.

During the tests, the samples' temperature evolution (thermocouples were placed on the samples' surface, center and bottom regions) and internal pressure (using sensors as proposed by Kalifa et al. [93]) were recorded. Additionally, such materials were dried with microwaves in the range of 2 to 6 kW/t at an atmospheric temperature of 120°C by feeding hot air into the device. Fig. 21 indicates the changes in temperature and internal pressure of the castable samples prepared with 5.6 wt.% and 6.3 wt.% of water.

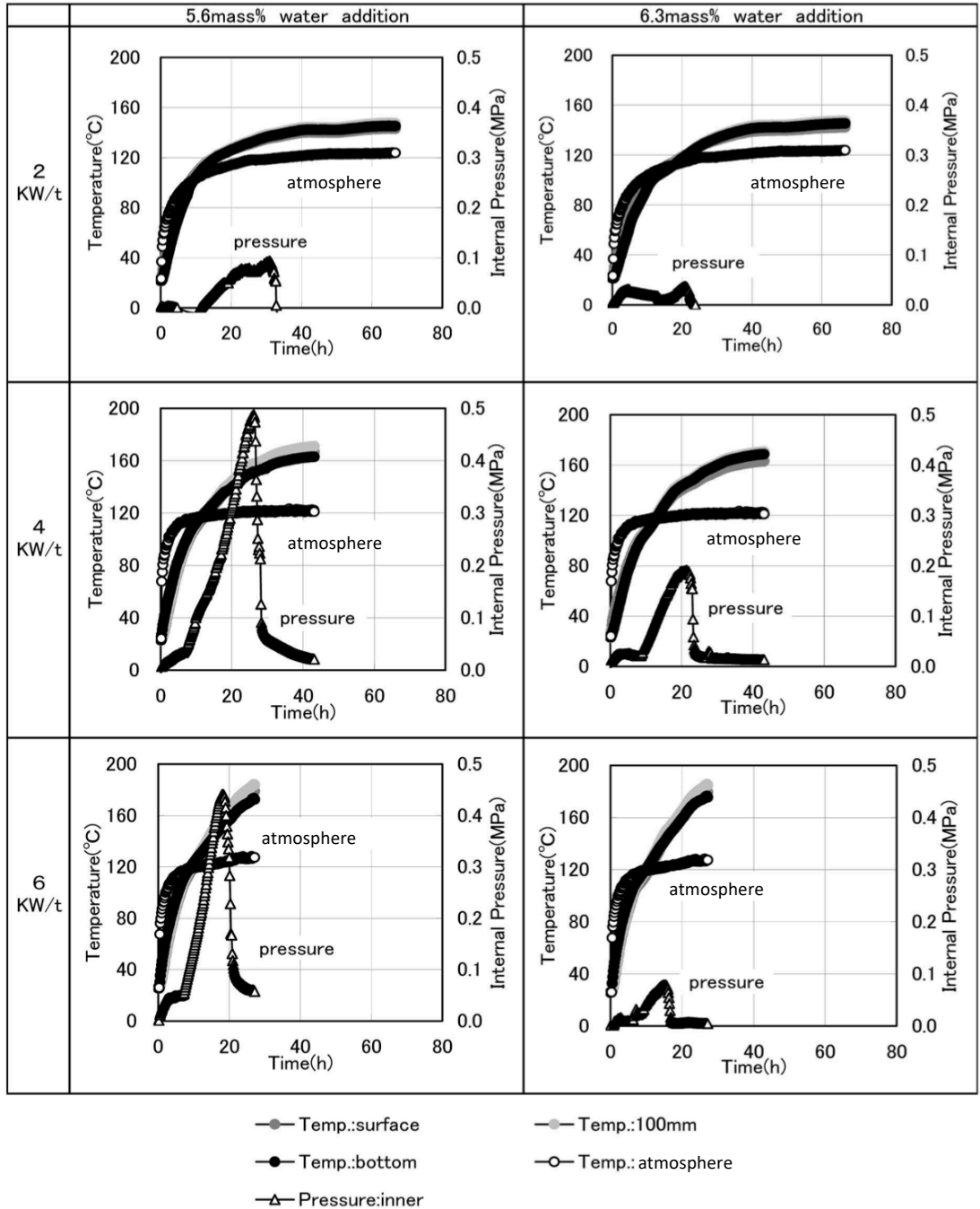

Figure 21: Temperature and internal pressure developed in the alumina-magnesia refractory castables during their drying step when subjected to the combination of microwave heating + hot air methods [92].

The entire samples were dried uniformly and minimal differences in the temperature profiles related to their surface, center and bottom regions were detected. Besides that, as the internal pressure is developed due to the steam formation, the time at which the measured values sharply decreased can be associated with those when the free water has been completely withdrawn (Fig. 21). Regarding the overall time required to dry the alumina-magnesia refractories subjected to different microwave output (from 2 to 6 kW/t), the increase of the latter allowed the reduction of the drying procedure from 32 h to 22 h for the mixtures prepared with 5.6 wt.% of water, whereas the ones containing 6.3 wt.% of liquid showed a time reduction from about 23 h to 17 h. Therefore, it was pointed out that using such technology might result in downtime savings of ladles and enable the reduction of the total number of furnaces required by steel plants.

### 2.3.3. *In situ* measurements of the water vapor pressure

Spalling of refractory castables is a complex phenomenon due to the coupling of the thermal, hydral, chemical and mechanical processes that take place during the heat up of these materials. These processes also compete with each other, as the generated flaws may help to increase the permeability to withdraw water and reduce the pore pressure; but it also decreases the castables' mechanical strength, increasing the spalling susceptibility.

Due to these aspects and aiming to measure the pore pressure evolution during the heat-up step of castables, some researchers have proposed various experimental setups [12,14,83,93–97]. As highlighted by Jansson [98], the most common ones used for the characterization of refractory castables and concretes designed for civil engineering applications are (Fig. 22):

- A. *Embedded pipe*, which transfers the pressure to an external pressure gauge. In this and other similar applications (B-D), the tube is cast inside the refractory and it exits the specimen away from the heat source.
- B. *Embedded pipe with internal rod*. The internal rod is used to reduce the hollow volume in the tube and a cavity is created around the measurement point.
- C. *Embedded pipe with clamped sintered porous material*. A thermocouple is included (inside or outside the tube) to record the temperature.
- D. *Embedded pressure gauge* for direct measurement of the pressure.
- E. *Embedded pipe*, in this case the tube is parallel to the heated surface.

For the measurement setup types A, B, C and E, the pressure is transferred to the outside of the castables using a medium (air, water, oil, mercury) inside the tube. Some authors [93,95] used air as a medium in the pipe, but when connecting it to the pressure gauge, a silicon oil filled flexible tube was also applied.

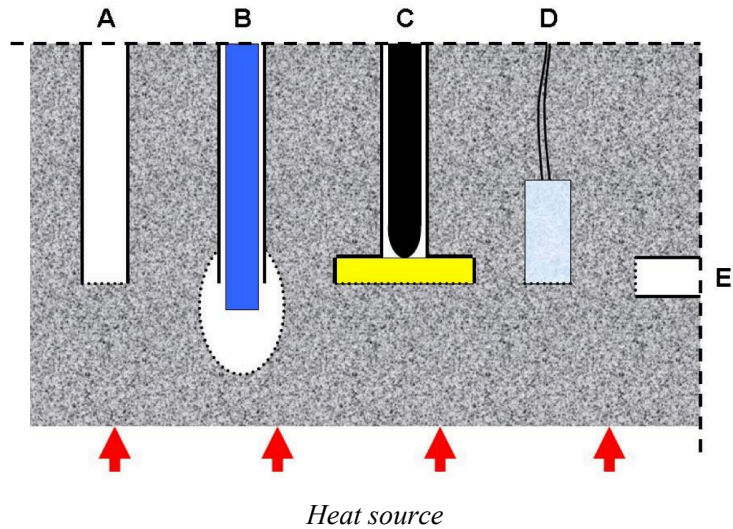

*Heat source*

Figure 22: Different pressure measurement setups (not to scale) applied in the spalling characterization of concrete materials [98].

Most of the papers published in the literature are based on setup C, as illustrated in Fig. 22. For instance, Kalifa et al. [93] carried out one-side heating experiments using prismatic specimens (300 mm x 300 mm x 120 mm) of ordinary (OC) and high-performance (HPC) concretes. The mass loss (due to evaporation of the liquid water) and the pore pressure of the tested materials were monitored by measuring the weight of the samples and using five pore pressure sensors (placed at 10, 20, 30, 40 and 50 mm from the heated surface of the samples), respectively, as shown in Fig. 23a. Each one of them consists of a sintered metal plate with controlled permeability, which is capsulated into a metal cup (impermeable material that acts supporting the metallic plate). The latter is brazed to a thin metal tube, which comes out of the rear face of the specimen. During testing, a tight connector was placed at the free end of the tube, linking the gauge to a piezoelectric pressure transducer (Fig. 23b). Additionally, a thermocouple was also inserted inside the tube to measure the temperature of the sintered metal plate. It is accepted that this setup (cup with sintered material) measures an average pressure over a larger area, whereas a bare pipe only provides a more local analysis, leading to a higher degree of variation of the results.

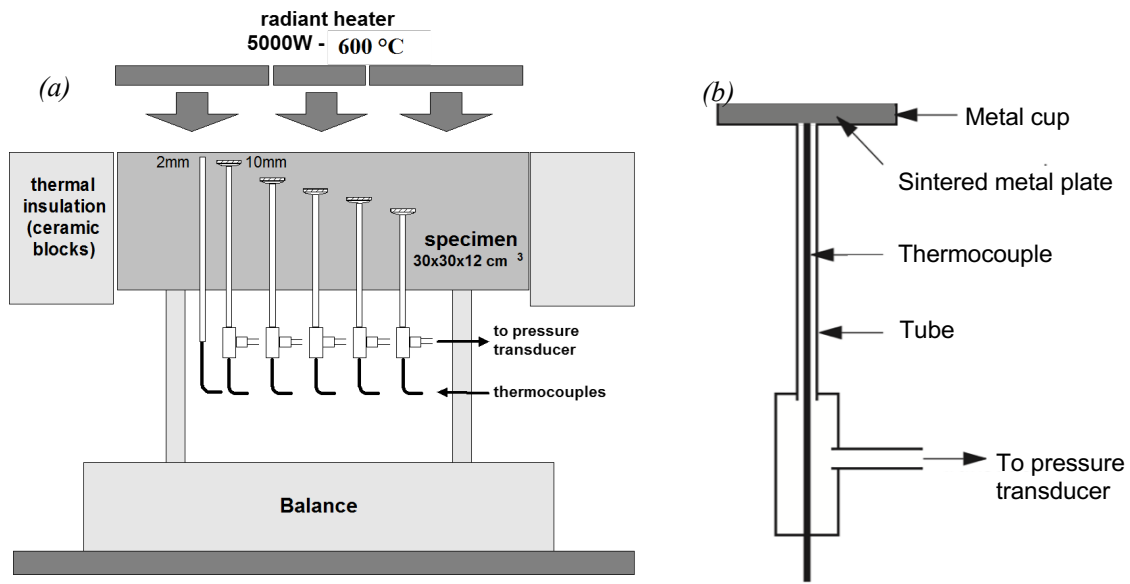

Figure 23: Pore pressure (a) experiment setup and (b) sensor used by Kalifa et at. [12,93].

Despite the fact that this sensor has been used by many researchers in other studies [6,14,83,95], some drawbacks may be pointed out. For instance, the high thermal conductivity of the metal tube conducts heat from the point of measurement to the cold zones of the sample, favoring the saturated vapor pressure's decrease. The condensation of the vapor phase contained in the tube as well as the re-evaporation of this liquified water (when the drying from moves inside the castable) should also influence the measured pressure. Another issue is that the metal cup is impermeable to vapor, air and liquid water, which inhibits these fluids to move through the measuring zone [12].

Meunier and colleagues [83] evaluated the pore pressure and temperature profiles of conventional (CC = 20 wt.% of CAC), medium cement (MCC = 10 wt.% CAC) and low cement (LCC = 8 wt.%) castables after their curing step at 20°C for 6 days. The same equipment and samples' shape illustrated in Fig. 23a were also used in those measurements and the sensors were placed at 2, 10, 20, 40, 60 and 90 mm from the refractories' hot face. Fig. 24 shows the temperature, mass loss and pressure evolution as a function of time at the different zones of the CC specimen. According to the results, as the heat reaches a given depth inside the sample, the pressure increases significantly due to water vaporization. Moreover, all tested materials presented their overall maximum pressure close to the central area of the specimens and the obtained values varied from 0.6 to 2.1 MPa, depending on the cement content and the resulting microstructure of the refractory [83].

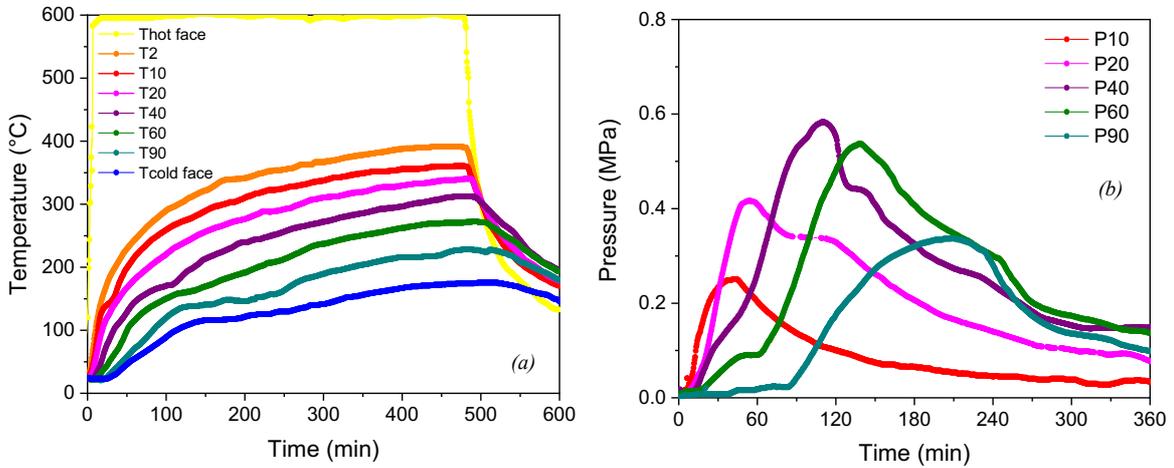

Figure 24: (a) Temperature and (b) pore pressure evolution as a function of time for conventional refractory castables (adapted from [83]).

The bell shape of the curves shown in Fig. 24b is related to the generation of a saturated zone between the drying front and the unexposed surface of the samples (the so-called moisture clog). As the vapor flow cannot go through this clog, it only moves in the opposite direction of the heat flux. Hence, as soon as the vapor rate leaving a pore is higher than the rate of vapor filling it (coming from vaporization, dehydration and transport), the pressure starts decreasing [95].

It was not mentioned by those authors [83] whether the experiments were reproducible or not. However, as highlighted by Mindeguia et al. [95], when using such measuring system for the evaluation of concrete specimens (for civil engineering application), the obtained pressure values were scattered even for two tests carried out in the same condition. As expected, these are local measurements and the pressure values depend on the tube location and their presence might modify the porous media around the metal plate. Additionally, the nonhomogeneous nature of the concrete structure could be responsible for the scattered results.

Aiming to mitigate some of these effects, Fey et al. [12] developed a new sensor also based on a porous sintered metal plate with a higher permeability ($k \sim 10^{-12}$ m$^2$) than the surrounding castable (Fig. 25a). In this case, the plate (40 mm x 40 mm x 2 mm) was placed inside the specimen (blocks of 300 mm x 300 mm x 120 mm cured at 20°C), parallel to the isotherms to prevent condensation. A thermocouple was placed inside the tube to measure the plate temperature and to minimize the hollow volume inside this device.

Fig. 25c highlights the pressure distribution developed inside a high-alumina refractory cured at 20°C and tested by Fey et al. [12]. The rising part of each curve (where S2 = 24 mm, S4 = 42 mm, S4 = 60 mm, S5 = 78 mm and S6 = 95 mm from the top surface of the block) presented only a small scattering as the obtained pressure values were very close to their equilibrium vapor pressure. However, higher scattering close to the peak profile (which reached pressures between 15-20 bar or 1.5-2.0 MPa) and the decreasing part of the curves was associated with the property variations of different specimens. These authors also observed that the highest pressure was detected in the center area of the blocks and that the measured values obtained at the colder face (S6) were higher than the ones of the hot surface (S2). The likely explanations for such behavior are: *(i)* two drying fronts will move contrawise in the x-direction; *(ii)* the reduced permeability of the cold zones; and *(iii)* the increasing amount of liquid water accumulated in the castable's cold side. Thus, these effects result in extra water availability to the second drying front, giving rise to higher pressures and slower motion of the heating front [12].

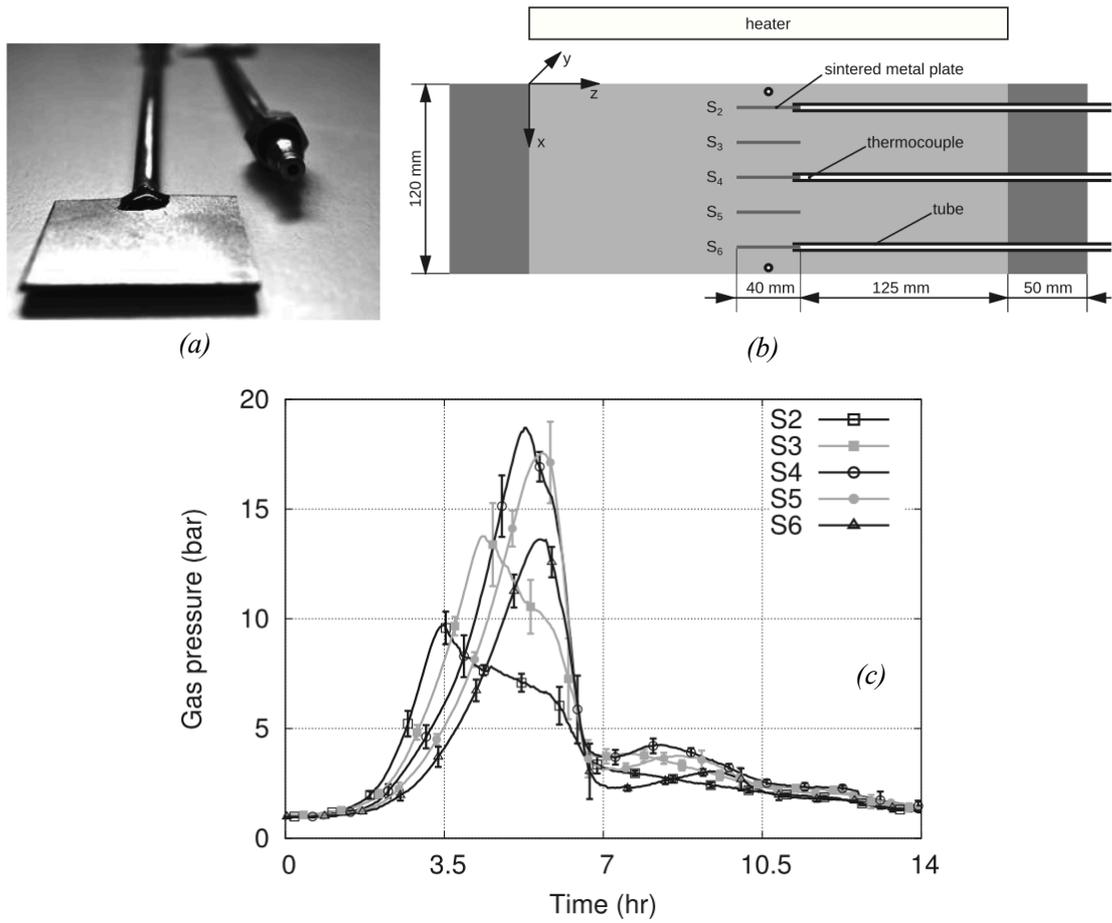

Figure 25: (a) Pressure sensor, (b) experimental setup (dark gray = insulating material; light gray = refractory castable) and (c) measured gas pressure versus time (average values and standard deviation) obtained for a refractory material tested by Fey et al. [12].

In general, the efforts for improving the quality of the sensors and collected pressure data are justified by the fact that the presence of such elements inside the castables' microstructure might change the samples´ properties (forming cracks or pores/voids near the probes) and affect the measurements. For example, the generated flaws around the sensor can potentially build an escape channel for vapor [99]. Moreover, the heterogeneous nature of castables' microstructures induce some randomness in the flow paths of vapor and might result in a non-uniform pressure distribution. Ultimately, it is accepted that a type D setup (see Fig. 22) with a very small pressure gauge and the electrical wires exiting the specimen along the isotherms would be preferable [98].

The type E setup, as the one used by Fey et al. [12], with a pipe exiting along the isotherms is also promising.

2.3.4. Magnetic resonance imaging (MRI), nuclear magnetic resonance (NMR) and X-ray computer tomography

In general, destructive and indirect methods are mainly used on a laboratory scale (such as TGA and DTA) to analyze the spalling likelihood of castables during heating. However, these techniques are limited, as they do not provide reliable data regarding the moisture distribution in the ceramic body and the evaluation of inadequate sample sizing might prevent the development of a temperature gradient in the tested specimens [100,101].

Thus, magnetic resonance imaging (MRI) and nuclear magnetic resonance (NMR) have been proposed as suitable options to follow the drying front movement inside ceramic samples [2,100–103]. MRI scanning works as an imaging method where the magnet of the scanner can act on the positively charged hydrogen ions ($H^+$) contained in the refractories and cause them to spin in an identical manner. Hence, by varying the strength and direction of this magnetic field, the spin direction of the protons might be changed, enabling one to build layers of the analyzed microstructure. Oummadi et al. [2], for instance, evaluated alumina and kaolin based samples (length = 40 mm, height and width = 15 mm) using the MRI technique, which provided information on the macroscopic water distribution in the green pieces with a spatial resolution finer than 0.5 mm. Such measurements were carried out with one side of the specimens exposed to drying at 40°C with room humidity. According to Fig. 26, for drying of 4h or more, a gradient in water concentration took place in the samples structure and, as the outer part of the green body becomes completely dry, the steepest part of the gradient (drying front) moves towards the inner part, as shown by the curve obtained after 52h [2].

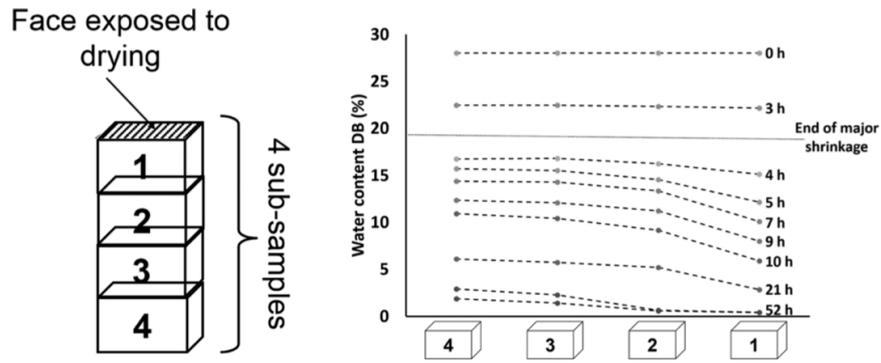

Figure 26: Water content in alumina samples after different drying times at 40°C and room relative humidity [2].

Nuclear magnetic resonance (NMR) is also a technique that can evaluate the hydrogen presence, and therefore the moisture profiles of green ceramic specimens can be nondestructively measured. As stated by Stelzner et al. [103], for the $^1$H-NMR measurement principle, a strong static magnetic field is used to study the relaxation of hydrogen nuclei (protons), available in abundance in water. The intrinsic atomic property of spin aligns with the static magnetic field and results in a small net magnetization, which can then be excited by radio-frequency (RF) and their relaxation back to the thermal equilibrium may be measured as an electromagnetic signal recorded by coils [103]. The initial amplitude of the measured electromagnetic signal (after the RF pulse) corresponds to the total amount of protons, and therefore the total amount of water within the ceramic can be estimated.

Barakat et al. [100–102] reported the development of a high-temperature nuclear magnetic resonance (NMR) setup, which was used for the characterization of regular (RC), low-cement (LCC) and non-cement (NCC) castables in the 100-300°C. As shown in Fig. 27, in such experiments the cylindrical sample (74 mm x 74 mm) was placed inside the birdcage coil and protected by an aluminum oxide ceramic tube. Thermocouples were placed inside small holes made in the tested specimen (7, 27, 44, 56, 64 and 71 mm from top to bottom) and this set was enclosed by a Faraday cage. A pair of anti-Helmholtz coils was used to generate the magnetic field gradient and an array of 100W halogen lamps provided the heat (heat flux = 3.5 kW/m$^2$ and heating rate ~2°C/min). In order to expose only one surface of the tested sample toward external heating, the cylinders were isolated with rock wool and Teflon to prevent heat and moisture loss.

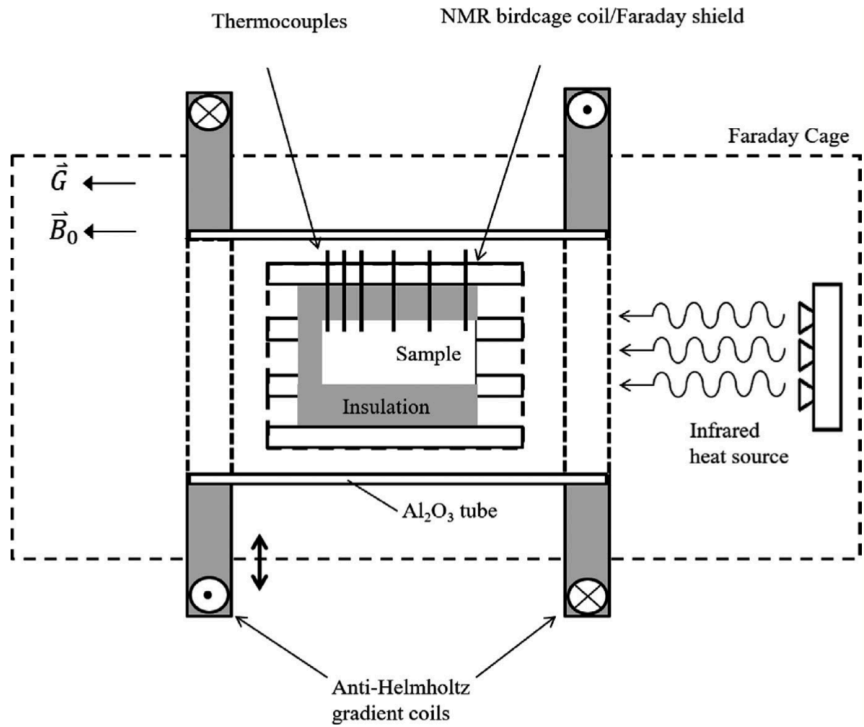

Figure 27: Schematic representation of the experimental NMR setup for measuring the moisture profiles during heating up of refractory castables [100].

Figure 28 shows the results collected during the NMR evaluation of a cured conventional castable (80 wt.% alumina and 20 wt.% CAC) [101]. The temperature profiles were measured simultaneously along with the saturation (moisture) profiles, with an average time of 3-4 min in between them, resulting in a total time of 1.5 h for each test. Based on the saturation profiles (Fig. 28a), an increase in the total moisture content was identified up to 20 min (Fig. 28b) due to water migration and phase transformations (i.e., conversion of metastable phases). On the other hand, the thermal gradient and beginning of the boiling front (~ 100°C, Fig. 28c and 28d) could be obtained by monitoring the temperature evolution in the refractory samples.

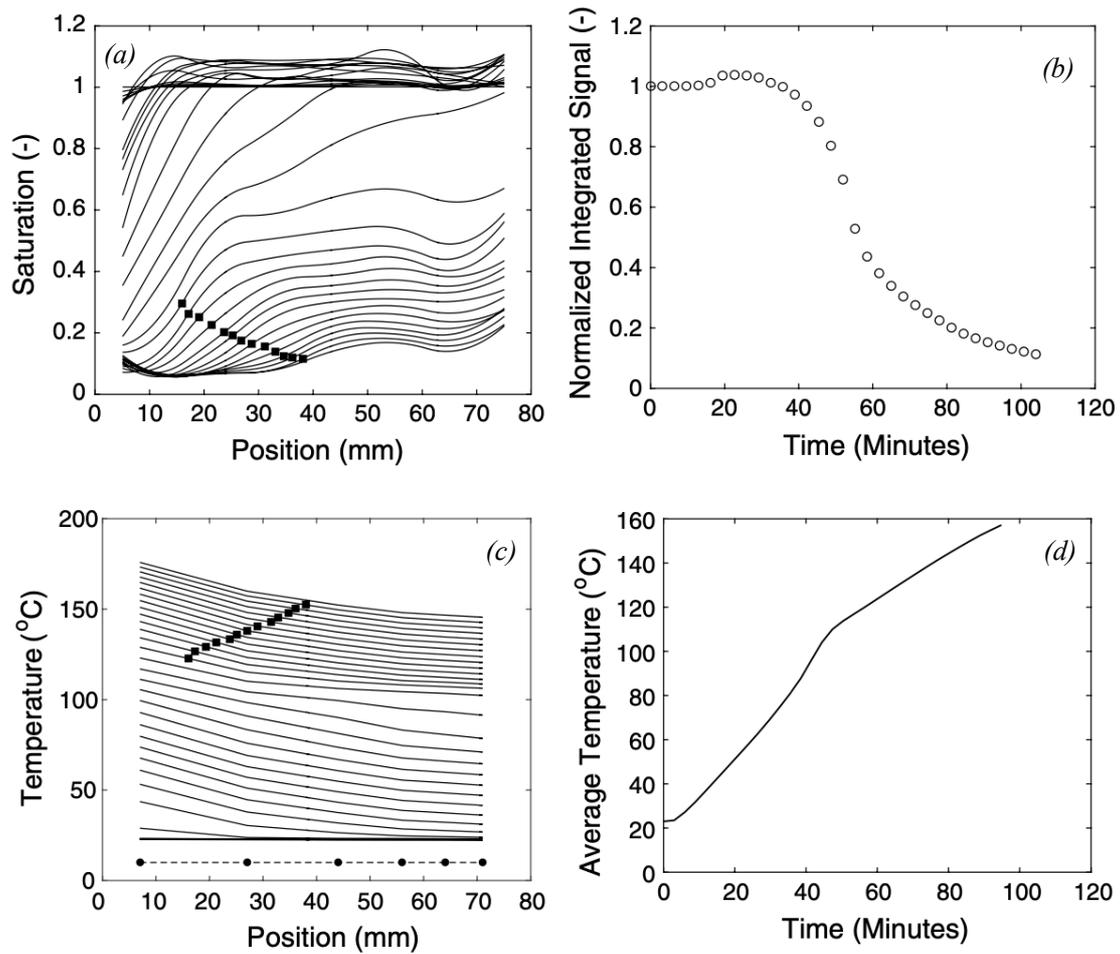

Figure 28: First drying of a conventional castable cured for 48h. (a) NMR saturation profiles, (b) total normalized saturation as a function of time, (c) temperature profiles and (d) average temperature with time. The black square markers in (a) and (c) are a visual guide for the drying front in the saturation and temperature profiles. The six circular markers in (c) correspond to the thermocouple positions [101].

After that, moisture was driven out very quickly until about 50 min, corresponding to the second drying stage and the water vaporization (Fig. 28b). These two stages of moisture loss were related to ambient conditions, such as temperature and humidity, which determined the drying speed. The black markers indicated in Fig. 28a and 28c highlight that the drying front (100-150°C) was developed after about 50% of the total moisture content was removed from the sample. While increasing the temperature, the samples' drying is then governed by internal transport processes that depends on pore structure and the refractory features, which makes this process slower than the previous stages [101]. This lower range of moisture content is usually considered as the most problematic from an industrial point of view, because it may lead to high steam pressure within the pores and, consequently, induce the castables' spalling.

The drying front position of cured castable samples can be identify via NMR measurements by taking the midpoint along the liquid-vapor boundary in the moisture distribution. Moreover, the critical saturation content corresponds to the transition point in the drying rate, from fast (externally limited) to slow (internally limited) drying. For instance, Fig. 29 shows the total average saturation content with time for four castables: RC (regular), LCC (low-cement), LCC Fume (low-cement containing silica fume) and NCC (non-cement). The plot of RC results points out the quantitative determination of the critical moisture (saturation) content by fitting two lines (solid lines) to the externally and internally limited drying stages. The intercept of these two segments defines the critical moisture content, which for RC is 0.4. For the other materials, the transition is less sharp due to their different characteristics (i.e., amount and type of hydrated phases, permeability level, etc.) [100].

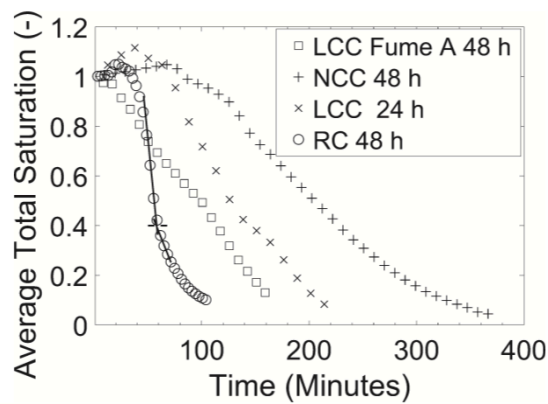

Figure 29: Average total saturation content as a function of time for four different materials: RC (regular), LCC (low-cement), LCC Fume (low-cement containing silica fume) and NCC (non-cement) castables [100].

The moisture transport and reconfiguration (determined by a change of the *in situ* bonding state of water) in high-performance concrete (HPC) were also investigated using X-ray-CT and $^1$H-NMR by Stelzner et al. [103]. Fig. 30 illustrates the devices used in this investigation.

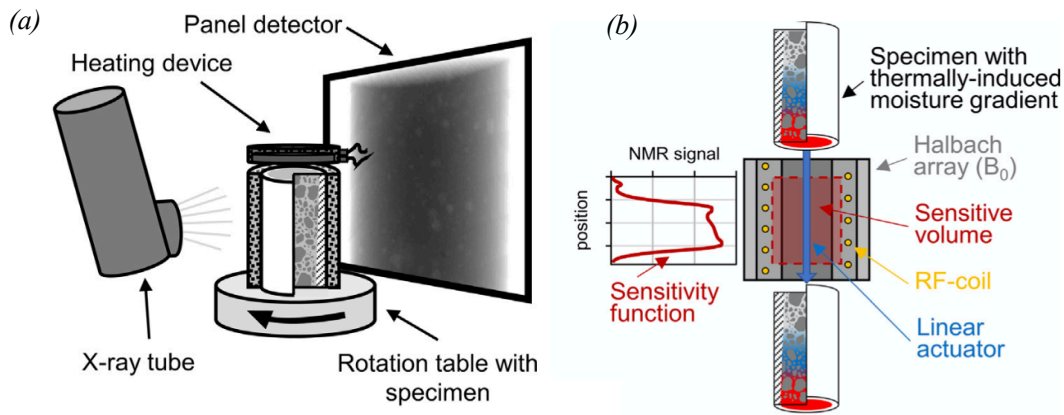

Figure 30: Sketch of (a) the HPC specimen and heating device inside the X-ray CT setup and (b) the movement of the specimen through the sensitive volume $^1$H-NMR measurements [103].

For the X-ray-CT, the signal attenuation correlates with the density of the material. Thus, changes in the water content can be visualized by the X-ray attenuation variation, as shown in Fig. 31. The moisture changes can be divided into the drying of the cement matrix (in red), which progresses from the heated side of the specimens; and the moisture accumulation (in blue) in deeper areas, including the saturation of the macro-pores with liquid water. Moreover, an increasing speed of the drying front inside the PP (polypropylene fibers)-containing samples could be observed due to the increased permeability of such concrete. By combining these results with the recorded temperature profiles for every time step, conclusions about the resulting pressure inside of the specimens might also be drawn, which makes X-ray-CT a very interesting technique due to the various pieces of information that can be obtained from these measurements [103].

The water content profiles for the same HPC compositions with and without PP-fibers were also evaluated by $^1$H-NMR and the collected data before and after the one-side heating are presented in Fig. 32. As observed, the interhydrate or free-water (that has not yet been involved in reactions) content was around 2 vol.% and almost homogeneously spatially distributed within the specimens before the heating step. The gel pore water (chemically-bonded one) content varied between 5 and 9 vol.%, showing higher values in the middle of the specimen than close to the specimen's surface, which was attributed to the differences in the hydration between the boundary area and the core of the sample or due to storage [103]. On the other hand, after the heating procedure (up to 300°C), the drying zone of the fiber-free concrete was located within the first 32 mm region, whereas the PP-containing ones showed a depth of 45 mm (Fig. 32).

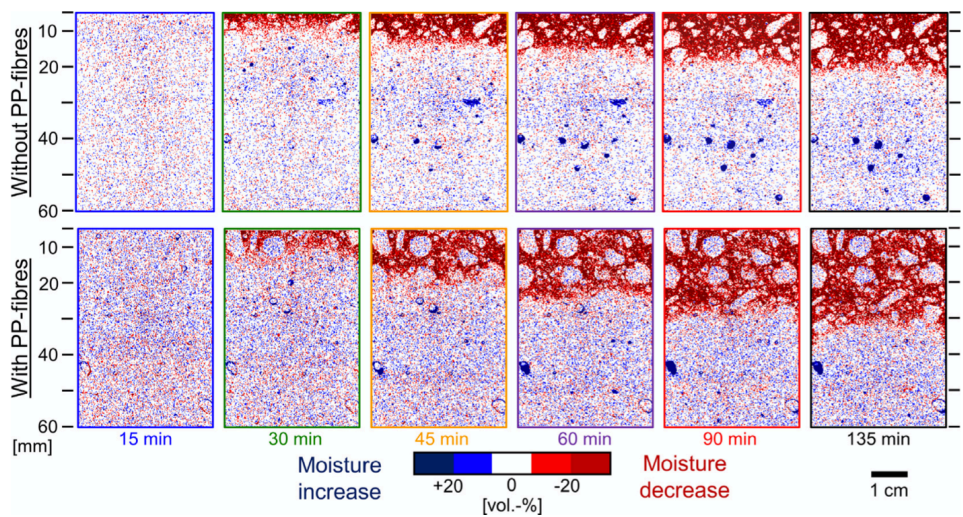

Figure 31: Moisture changes inside the high-performance concrete with and without PP (polypropylene)-fibers during unilateral heating at different time steps measured using X-ray-CT [103].

According to Fig. 32, a maximum moisture content increase within the tested samples was around 4 vol.%, which was followed by a greater amount of moisture located in larger pores (free-water) and reduction of the chemically-bonded (gel) water. This indicates that a reconfiguration of the moisture from small pores (gel) to larger ones (interhydrate) took place during heating. Besides that, the total moisture content at the bottom of the specimen increased about 5 vol.% and 2 vol.% for the fiber-free and PP-containing concretes, respectively, after heating [103].

Based on the examples presented above, MRI and NMR are methods that can provide additional information and complement the traditional techniques used to characterize the drying behavior of dense materials. Collecting reliable data on the moisture distribution throughout the first drying of refractories will help to narrow the assumptions commonly made in numerical modeling regarding the internal transport dynamics in ceramics. As a research tool, NMR is also capable of directly quantifying local features of drying, which allows preserving sample dimensions (non-destructive) and providing direct insight into physical phenomena resulting from fiber burn-out, vapor release, etc. [102]. Hence, it is expected that more efforts should be placed in trying to apply such techniques to investigate the drying behavior of different refractory castable systems.

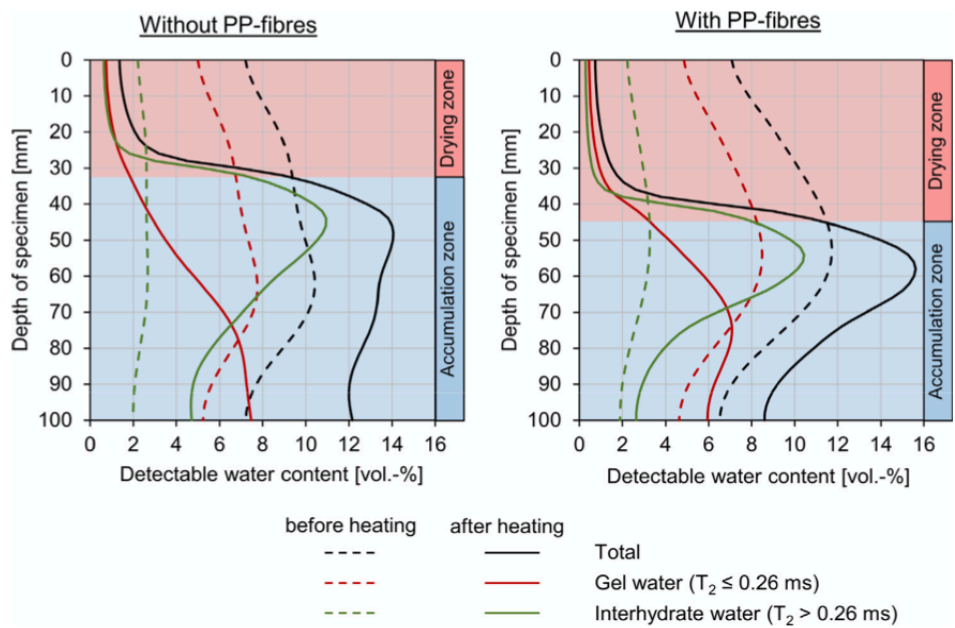

Figure 32: Comparison of the spatial water content distribution detected by 1H-NMR before and after the one-side heating of HPC specimens with and without PP (polypropylene)-fibers [103]

2.3.5. Neutron tomography

Spalling consists of an intrinsically three-dimensional process as it locally depends on the heterogeneity of the microstructure. Hence, to capture the complex mechanisms in their entirety involved in such a phenomenon, the evaluation of 3D moisture profiles based on neutron tomography seems to be more appropriate than the one-dimensional piece of information provided by NMR [100,101] and neutron radiography [104].

As reported by Dauti et al. [27,105,106], neutron tomography is an innovative method which provides access to the local moisture distribution. Its principle is based on the attenuation, through both scattering and absorption, of a directional neutron beam by the matter through which it crosses. Because hydrogen atoms have the ability to absorb a significant magnitude of this beam, the particularly high attenuation resulting from this process is especially convenient to study any drying fronts in dense ceramics, as illustrated in Fig. 33.

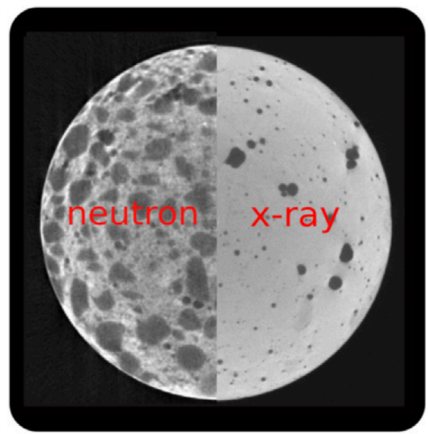

Figure 33: Comparison of two horizontal slices of neutron and X-ray tomography images of the same cylindrical sample of concrete (70 mm in diameter) acquired in comparable conditions [105].

When comparing neutron and X-ray tomography images of the same cylindrical concrete sample (Fig. 33), the latter is ideal to detect pores (because of their low density) but cannot distinguish the fine components of the matrix' fraction and the coarse grains due to their comparable density and atomic number. Conversely, the aggregates are almost invisible to the neutrons, which makes their distinction from the matrix region straightforward but hinders their differentiation from the pores. This feature enabled us to investigate the moisture content evolution in the cement paste contained in concretes [105]. Although this technique has not been applied yet to investigate the drying behavior of refractory castables (Dauti et al. [27,105] were pioneers in presenting 3D *in situ* neutron tomographies of concretes), the following examples are presented in order to highlight the interesting data that can be obtained in such measurements.

Neutrons can penetrate concrete only through a few centimeters. Thus, this technique is only applicable to small samples (i.e., cylinders with a diameter of 3-7 cm and length of 3-5 cm). Fig. 34 illustrates the 3D volume images of high-performance concretes containing aggregates (mixture of sandstone, dolomite and metamorphic rocks) with maximum size of 4 mm or 8 mm. Due to the strong attenuation of the wet cement paste, the evolution of the drying front is evident in the images. However, despite using heat and moisture insulating materials (aluminum tape and rockwool) around the lateral surface of the samples, drying also progressed at the boundaries. Thus, the presented results show qualitatively how the drying front moves faster in compositions containing bigger aggregates (Fig. 34). Nevertheless, image processing can be used to overcome the qualitative nature of these observations, as aggregates and cement paste can be separated using thresholding techniques [27].

For instance, white pixels might be assigned to the aggregates and the black ones to the cement paste (Fig. 35). As the drying of the samples takes place, and the water contained in the cement paste migrates, the contrast between this region and the coarse grains reduces, hindering their distinction. However, the spatial distribution of the aggregates (measured at the initial state), can be assumed not to change throughout the test. Consequently, it is possible to subtract them from every tomographic picture, leaving behind only the cement paste, which can be in a dry and/or wet state. From this point, it is straightforward to isolate through thresholding the dry and wet parts of the sample (Fig. 35b).

Fig. 35c points out the ratio of the dehydrated volume ($V_{dry}/V_{initial}$) with the time of the wet HPC concretes, where it can be observed how the sample with bigger aggregates dried faster. Drying started after about 10 min for both concretes when the temperature at 3 mm from the surface was around 170°C. Moreover, the gray region close to the blue lines represents the uncertainty induced by choice of the threshold when separating the wet and dry cement [27].

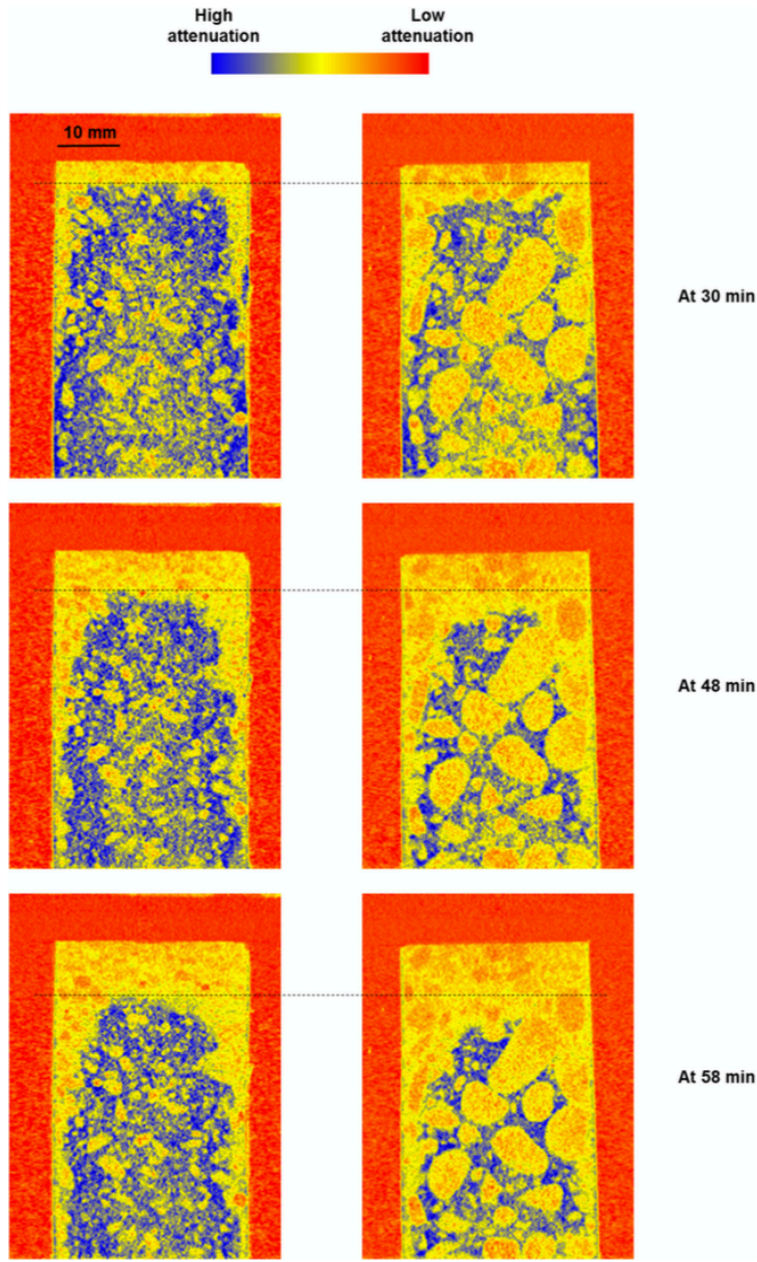

Figure 34: Vertical slices from reconstructed 3D volumes showing the evolution of the drying front at different heating times in high performance concrete (HPC) 4 mm (on the left) and 8 mm (on the right). The samples were stored in plastic bags with 97% relative humidity and at 20°C before testing. [105].

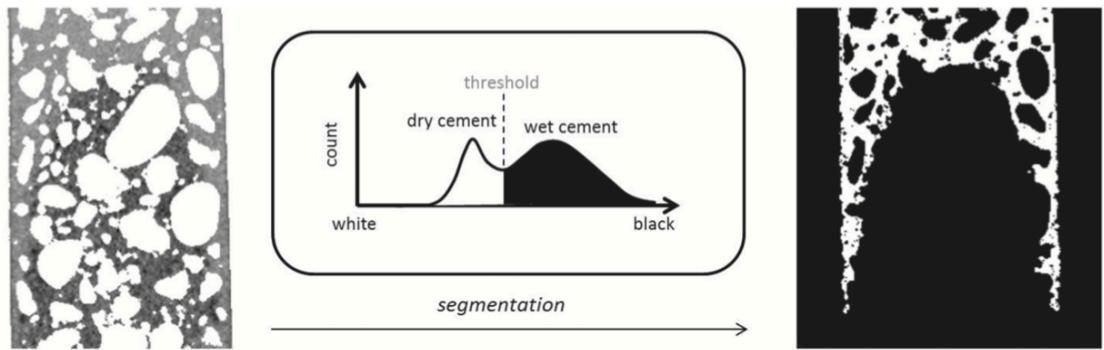

*(a)* *(b)*

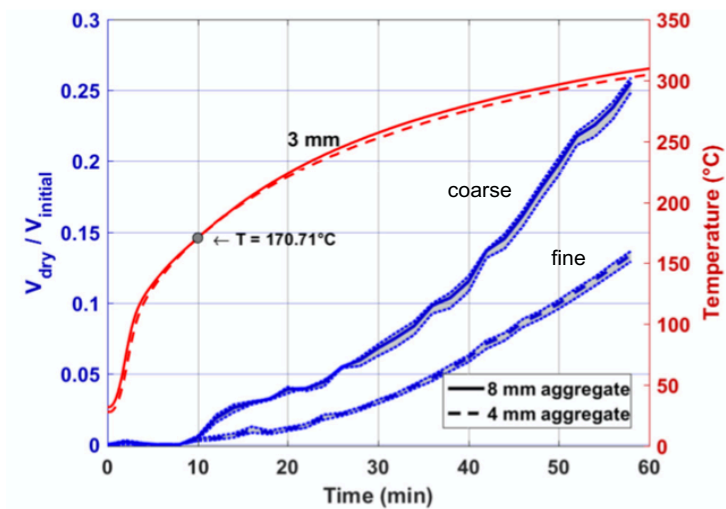

*(c)*

Figure 35: Quantification of the drying process through segmentation. (a) The aggregates obtained in the initial state are removed, leaving behind only dry and wet cement. (b) These two phases can then be separated through segmentation. (c) Evolution of the drying front with time when only the core part of the samples was considered [27].

Perhaps the most important input of the neutron tomography data presented by Dauti et al. [27,105] is the direct proof and quantification of the so-called "moisture clog" in the tested concrete samples. This accumulation of water behind the drying front could be inferred by analyzing the attenuation coefficient across the transverse section (xy-plane) of the samples and to plot this value as a function of the depth from the exposed face (z-position) and the elapsed time (Fig. 36a). Thus, the contrast between dried and wet zones allows tracking in time and space the drying front, as well as the end of the apparent moisture accumulation zone, which can be computed by finding the transition point in terms of averaged gray value. The blue shaded area, shown in Fig. 36b and 36c, represents the depth of moisture accumulation zone for the concretes with aggregates size of 8 mm and 4 mm, respectively.

The vapor produced at the drying front that moves ahead and condensates in the colder region of the dense ceramic structure (so-called "moisture clog") is pointed out as one of the main causes for the pressure build-up in such materials during heating. While some numerical simulations [107] corroborate this effect, limited experimental observations and quantification have been carried out so far. Therefore, the neutron tomography technique might be an important tool to be applied in the study of such a phenomenon.

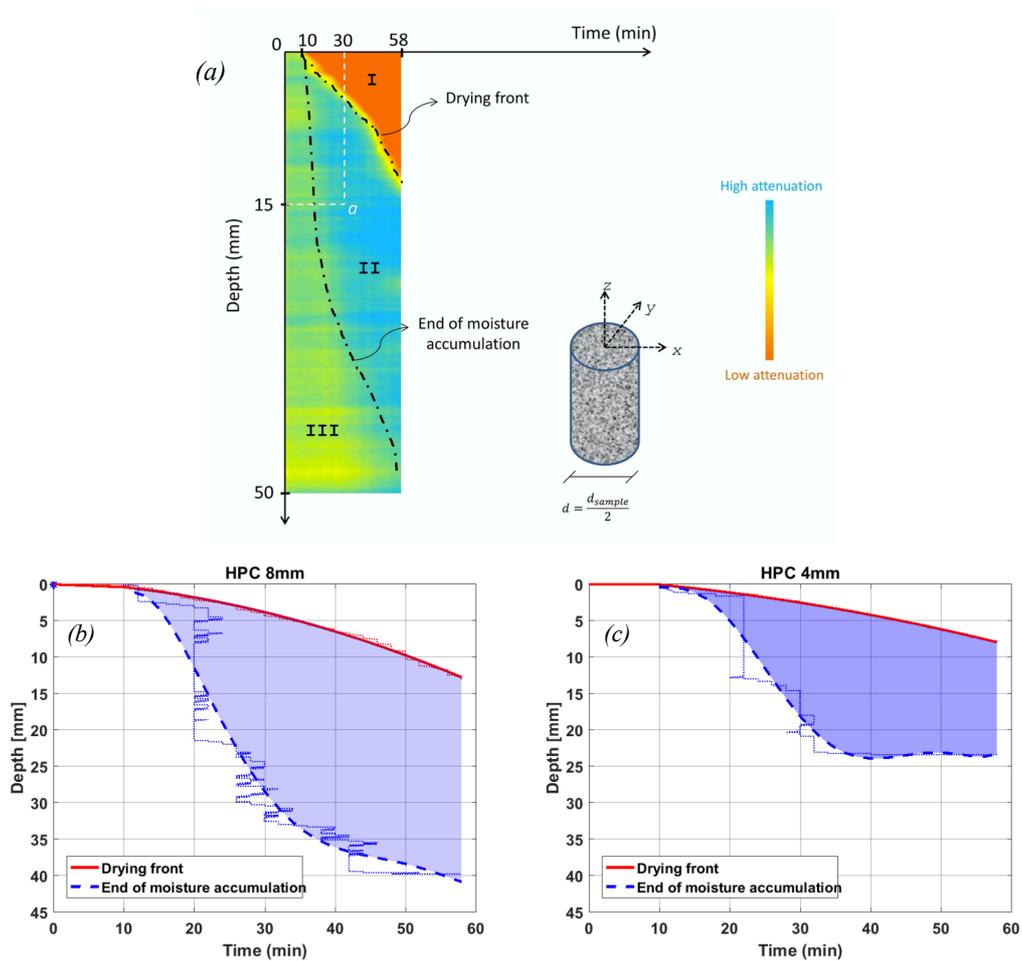

Figure 36: (a) Sketch of the moisture profiles for HPC containing aggregates with 8 mm. Each value in the graph (scaled by the color bar) represents the average attenuation in xy-plane. Depth of moisture accumulation zone for the concrete samples containing aggregates with size of (b) 8 mm and (c) 4 mm [105].

## 3. Final remarks

The search for advanced refractories highlighted one of the biggest challenges for monolithic applications, which is the design of proper drying schedules to ensure safe and fast heating of the ceramic linings. Drying is a time-demanding process, and therefore not only the costs related to the energy consumption, but also the impacts on the production halt (lost income) need to be taken into account. However, in industrial practice, the selection of appropriate heating rates and dwell times is commonly based on empirical knowledge.

This review presented the main drying steps of dense refractories and provided some insights about the optimization of the applied schedules (by inducing higher water withdrawal during the evaporation period, decreasing the amount of liquid to be released in the ebullition one). Thus, the phase transformations associated with different binder additives were discussed, as the heating procedure should be design according to the generated hydrated phases and their respective decomposition temperatures.

Despite the extensive research on determining the different factors influencing spalling and measures to avoid it, the mechanisms behind this phenomenon are not yet fully understood. Different methods are used for refractory spalling assessment. Although the large-scale tests are the most reliable, they are very costly. Thus, experimental techniques such as "macro"-TGA, *in situ* pressure measurements, high temperature NMR, neutron tomography and others are important options to be applied for the investigation of the drying behavior of refractory castables. As discussed in the previous sections, the methods that consider embedded elements, such as thermocouples and pressure gauges, need to be used with caution, as voids and flaws can be formed around them, affecting the evaluated parameters. This is particularly important when the pressure inside the castables is monitored, as the validation of many state-of-the-art numerical models on heated refractory is based on such measurements. Therefore, many efforts have been carried out to study the application of in situ imaging techniques (such as NMR and neutron tomography) to register the moisture evolution during the heating of dense ceramics. Some preliminary tests involving the analyses of refractory castables via neutron tomography led to some promising results, which might result in new insights about the water removal of such ceramics.

Besides these aspects, a forthcoming review paper by the authors (Part II) should discuss the main drying steps and provide guidance to deal with them without damaging the castable's structural integrity. Thus, the following subjects will be presented: *(i)* main drying agents and how they affect the resulting refractories' microstructure (organic fibers, metallic powders, permeability enhancing active compounds, silica-based additives and chelating agents), and *(ii)* the design of drying schedules for different refractory compositions (i.e., the role of heating rate, ramp versus holding time routine, and castable's dimension on the overall drying behavior).

## 4. Acknowledgments

The authors would like to thank the Conselho Nacional de Desenvolvimento Cientifico e Tecnologica – CNPq (grant number: 303324/2019-8) and the Fundação de Amparo a Pesquisa do Estado de São Paulo – FAPESP (grant number: 2019/07996-0) for supporting this work.